\UseRawInputEncoding
\documentclass[journal=jpcafh,manuscript=article]{achemso}
\usepackage{longtable}
\usepackage{multirow}
\usepackage{amsmath}
\usepackage{subfigure}

\usepackage[usenames,dvipsnames]{color}
\usepackage{soul}
\usepackage{bm}
\usepackage{rotating}

\usepackage{amssymb}
\usepackage{titlecaps}
\Addlcwords{a an the of on at with for and in into from by without}

\makeatletter
\newcommand*{\addFileDependency}[1]{
  \typeout{(#1)}
  \@addtofilelist{#1}
  \IfFileExists{#1}{}{\typeout{No file #1.}}
}
\makeatother

\newcommand*{\myexternaldocument}[1]{%
    \externaldocument{#1}%
    \addFileDependency{#1.tex}%
    \addFileDependency{#1.aux}%
}
\usepackage{xr}
\myexternaldocument{si}

\usepackage{multicol} \usepackage{wrapfig} \usepackage{enumitem}
\usepackage{bm} \usepackage{outlines} %for itemize with subitems
\SectionNumbersOn

\author{Meenu Upadhyay}
\affiliation[University of Basel]{Department of Chemistry, University
  of Basel, Klingelbergstrasse 80 , CH-4056 Basel, Switzerland.}
\author{Markus Meuwly}
\affiliation[University of Basel]{Department of Chemistry, University
  of Basel, Klingelbergstrasse 80 , CH-4056 Basel, Switzerland.}
\email{m.meuwly@unibas.ch}

\title{Thermal and Vibrationally Activated Decomposition of the
  syn-CH$_3$CHOO Criegee Intermediate}

\begin{document}
\date{\today}

\begin{abstract}
The full reaction pathway between the syn-CH$_3$CHOO Criegee
Intermediate via vinyl hydroxyperoxide to OH+CH$_2$COH is followed for
vibrationally excited and thermally prepared reactants. The rates from
vibrational excitation are consistent with those found from
experiments and tunneling is not required for reactivity at all
initial conditions probed. For vibrationally excited reactant, VHP
accumulates and becomes a bottleneck for the reaction. The two
preparations - relevant for laboratory studies and conditions in the
atmosphere - lead to a difference of close to one order of magnitude
in OH production ($\sim 5$ \% vs. 35 \%) on the 1 ns time scale which
is an important determinant for the chemical evolution of the
atmosphere.
\end{abstract}

\noindent
The hydroxyl radical (OH), one of the most powerful oxidizing agents,
plays an important role in the chemical evolution of the
atmosphere.\cite{ stone:2012} OH, also referred to as the ``detergent
of the
troposphere''\cite{gligorovski2015environmental,levy1971normal},
triggers degradation of many pollutants including volatile organic
compounds (VOCs) and is an important chain initiator in most oxidation
processes in the atmosphere. The amount of OH generated from alkene
ozonolysis is an important determinant required for chemical models of
the lower atmosphere. Field studies have suggested that ozonolysis of
alkenes is responsible for the production of about one third of the
atmospheric OH radicals during daytime, and is the predominant source
of OH radicals at night.\cite{emmerson2009night,khan2018criegee}
Alkene ozonolysis proceeds through a 1,3-cycloaddition of ozone across
the C=C bond to form a primary ozonide which then decomposes into
carbonyl compounds and energized carbonyl oxides, known as Criegee
Intermediates (CIs)\cite{criegee1949ozonisierung}. These highly
energized intermediates rapidly undergo either unimolecular decay to
hydroxyl radicals\cite{alam2011total} or collisional
stabilization\cite{novelli2014direct}. Stabilized Criegee
intermediates can isomerize and decompose into products including the
OH radical, or undergo bimolecular reactions with water vapor, SO$_2$,
NO$_2$ and acids\cite{taatjes2017criegee,mauldin2012new}. The high
energy and short lifetime of these zwitterionic species complicates
their direct experimental characterization.\\

\noindent
One of the smallest CIs that can either follow unimolecular decay to
generate OH or bimolecular reactions under atmospheric conditions is
the acetaldehyde oxide (CH$_3$CHOO). This species is generated from
ozonolysis of trans-2-butene.\cite{atkinson:1993} Unimolecular
decomposition of stabilized \textit{syn}-CH$_3$CHOO proceeds through a
five membered transition state with an energy barrier of $\sim 17$
kcal/mol\cite{kuwata:2010, liu:2014,vereecken2017unimolecular},
following 1,4- hydrogen transfer to form vinyl hydroperoxide
(VHP). Subsequent homolytic cleavage of the O$_{\rm A}$--O$_{\rm B}$H
bond (see Figure \ref{fig:pes}A) leads to OH and vinoxy
radical.\cite{gutbrod:1997} Conversely, starting from the
\textit{anti}-CH$_3$CHOO isomer, the main product is methyl-dioxirane
which proceeds through a ring-closure step.\cite{long2016atmospheric}
The conversion of the \textit{syn}- to the \textit{anti}-isomer
involves a barrier of $\sim 42$ kcal/mol which makes such a
reorganization highly unlikely.\cite{yin2017does}\\

\noindent
Direct time-domain experimental rates, for appearance of OH from
unimolecular dissociation of \textit{syn}-CH$_3$CHOO under collision
free conditions were obtained by vibrationally activating the
molecules at specific energies in the vicinity of\cite{fang:2016} and
below\cite{fang:2016deep} the transition state barrier of 1,4 hydrogen
transfer. Statistical RRKM\cite{baer1996unimolecular} rates with
tunneling and zero point energy correction agreed with experimentally
determined OH formation rates at energies in the vicinity of the
barrier ($\sim$ 6000 cm$^{-1}$). Experiments on deuterated
\textit{syn}-CD$_3$CHOO found a kinetic isotope effect of $\sim 50$
for unimolecular decay at energies near the barrier which suggests
that tunneling occurs.\cite{green2017selective} Later computational
work reported the possibility for rearrangement of the vinoxy and
hydroxyl radical to form hydroxyacetaldehyde instead of O-O bond
homolysis in VHP.\cite{kuwata:2018} The dynamics of energized
\textit{syn}-CH$_3$CHOO from simulations initiated at the TS towards
VHP reported prompt OH-dissociation without visiting VHP along the
pathway.\cite{wang:2016two} Finally, thermal unimolecular decay is
also a relevant pathway for \textit{syn}-CH$_3$CHOO loss in the
atmosphere with rates at room temperature of 100 to 300
s$^{-1}$.\cite{nguyen:2016communication,zhou:2019}\\

\noindent
The second step for OH formation is O--O cleavage starting from VHP
which converts singlet VHP into two doublet radicals (OH and
CH$_2$COH).\cite{kurten:2012} At the MP2 level of theory reported
electronic energy differences between VHP and the dissociation
products of 31.5 kcal/mol and 35.8 kcal/mol using the 6-31G(d) and
aug-cc-pVTZ basis sets, respectively.\cite{kurten:2012} Calculations
at the multi-reference configuration interaction singles and doubles
(MRCISD) level of theory report dissociation energies between 14.3 and
17.8 kcal/mol depending on the basis set used. However, as the
asymptotic energy differences appear to depend both, on the size of
the active space and the basis set used, a definitive dissociation
energy for this process is currently not available. In another, more
recent effort\cite{lester:2016} the PES for the OH elimination
reaction was determined at the CASPT2(12,10)/cc-pVDZ level of theory
which find a first submerged barrier, 23.0 kcal/mol above VHP, and the
second, ``positive barrier'', 29.3 kcal/mol above VHP. Both of them
connect to the same asymptotic state (CH$_2$COH+OH) which is 27.2
kcal/mol above the VHP energy.\\

\noindent
For the \textit{syn}-CH$_3$CHOO $\rightarrow$ VHP $\rightarrow$
CH$_2$COH + OH reaction it has been suggested that the first step -
1,4-hydrogen shift - is rate limiting, followed by a rapid homolysis
of the O--O bond.\cite{marston:2008,gutbrod:1997,kurten:2012} On the
other hand, there has also been evidence that VHP itself is a
significant bottleneck along the reaction
coordinate.\cite{donahue:2011} This view is supported by the notion
that with an estimated O--O bond energy of $\sim 19$ kcal/mol thermal
decomposition of VHP would require a unimolecular rate below $10^9$
s$^{-1}$ which is not consistent with simple
scission.\cite{donahue:2011} Hence, there is also uncertainty on the
question which - if any - of the two steps is clearly rate limiting.\\

\noindent
For a more atomistically resolved understanding of the entire pathway
\textit{syn}-CH$_3$CHOO $\rightarrow$ CH$_2$CHOOH $\rightarrow$
CH$_2$CHO+OH, a statistically significant number of reactive
trajectories based on full-dimensional reactive potential energy
surfaces is run and analyzed in the present work. For this, reactive
MD simulations are carried out based on an empirical multisurface
adiabatic reactive PES (using CHARMM\cite{brooks2009charmm}) and a
neural network-based PES using the atomic simulation environment
(ASE)\cite{larsen2017atomic} for comparison.\\

\section*{Results and Discussion}
\begin{figure}[h!]
    \centering \includegraphics[scale=0.37]{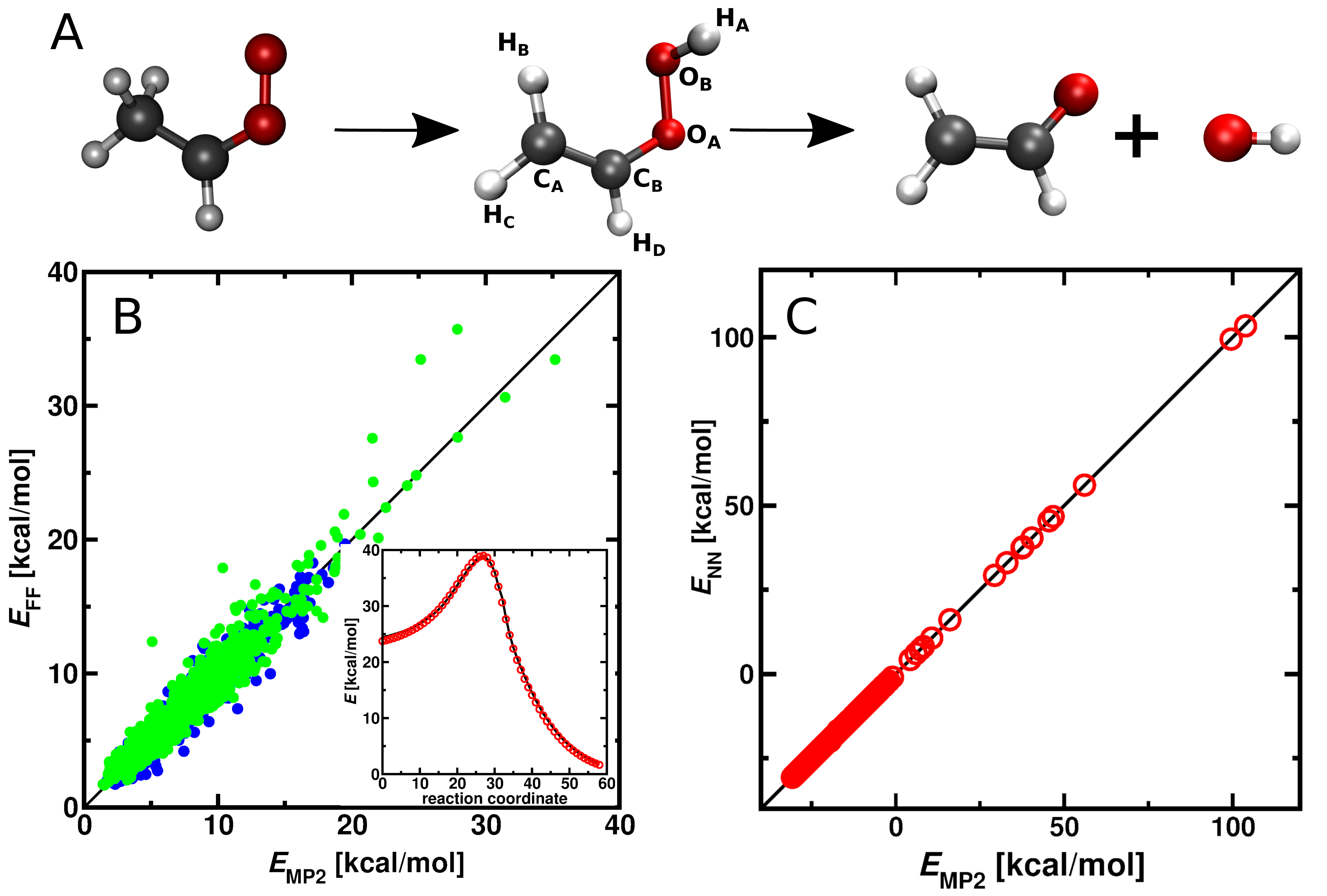}
    \caption{Quality of the PESs. Panel A: Schematic representation of
      OH formation starting from the \textit{syn}-CH$_3$CHOO Criegee
      intermediate (left) via vinyl hydroperoxide (VHP, middle)
      through a 1,4 hydrogen shift reaction and subsequent
      OH-elimination to yield vinoxy radical + OH (right). Panel B:
      Correlation of 1612 \textit{ab initio} reference structures and
      the fitted force field for reactant (blue) and product (green)
      with RMSE value of 1.1 kcal/mol and 1.2 kcal/mol
      respectively. Inset: \textit{ab initio} IRC (red circles) and
      fitted MS-ARMD (black curve). Panel C: Correlation between
      reference (MP2) and predicted (PhysNet) energies. The 10540 test
      set structures are predicted with MAE of 0.02 kcal/mol and RMSE
      of 0.19 kcal/mol.}
    \label{fig:pes}
\end{figure}

{\bf The Interaction Potentials} The quality of the MS-ARMD and
PhysNet representations of the reactive PESs is reported in Figures
\ref{fig:pes}B and C. For MS-ARMD the fitted PESs for the reactant
(blue, CH$_3$CHOO) and the product (green, VHP) states have overall
root mean squared errors (RMSEs) of 1.1 kcal/mol and 1.2 kcal/mol,
typical for such an
approach.\cite{MM.n4:2012,reyes.pccp.2014.msarmd,MM.mgo:2020} The IRC
closely follows the reference MP2 calculations, see inset Figure
\ref{fig:pes}B, which underlines the quality of the reactive PES. For
OH-elimination a conventional Morse fit to the reference MP2
calculations was used with $r_e^{\rm OO}=1.45$ \AA\/, $D_0^{\rm OO} =
31.5$ kcal/mol, and $\beta = 2.3$ \AA\/$^{-1}$. However, because O--O
bond breaking may involve multi-reference
character\cite{kurten:2012,lester:2016} simulations with a value of
$D_e^{\rm OO} = 23.5$ kcal/mol were also carried out, consistent with
findings from CCSD(T)-F12b/CASPT2 calculations.\cite{lester:2016}\\

\noindent
The quality of the PhysNet representation of the global reactive PES
is reported in Figure \ref{fig:pes}C. Here, the mean average error on
the test set ($\sim 10000$ structures not used for training or
validation) is 0.02 kcal/mol with a RMSE of 0.19 kcal/mol and $R^2 =
1-10^{-7}$ . Again, this performance is in line with previous
work.\cite{MM.atmos:2020,MM.ht:2020,MM.diels:2019}\\

\noindent
As a first exploration of the PES the minimum dynamic path (MDP) for
the 1,4 hydrogen shift reaction was determined on the PhysNet PES, see
Figure S2. Starting from the initial structure ``A'',
this reaction passes through a five-membered ring (structure ``C'')
before formation of VHP (structure ``E'').\\

\begin{figure}
    \centering \includegraphics[scale=0.45]{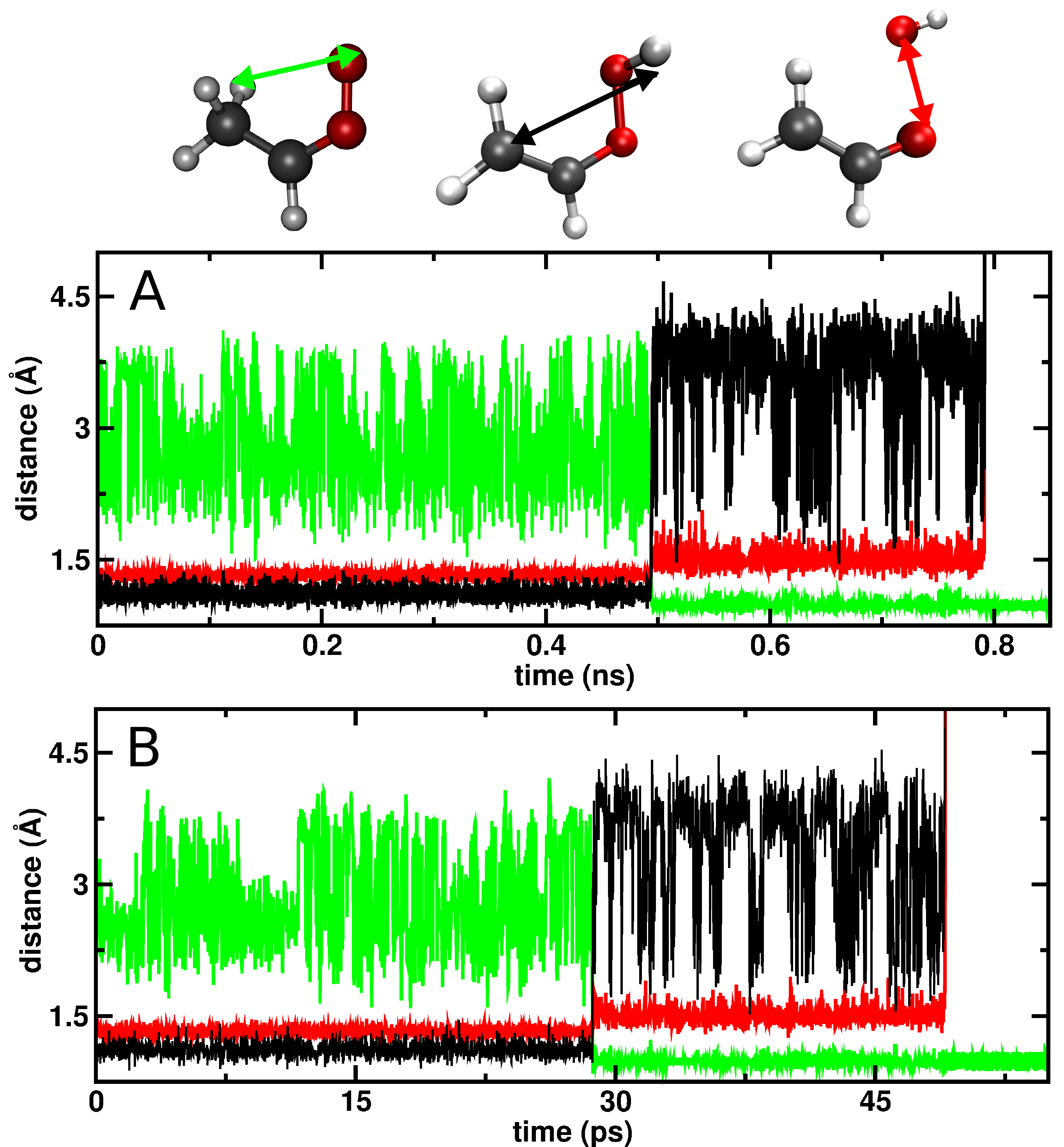}
    \caption{Time series for different distances for the reaction
      \textit{syn}-CH$_3$CHOO $\rightarrow$ VHP $\rightarrow$
      CH$_2$CHO+OH. Panel A: using MS-ARMD simulations at $t \sim 0.5$
      ns VHP forms and at $t \sim 0.79$ ns OH elimination takes
      place. Panel B: using PhysNet at $t \sim 28$ ps VHP forms and at
      $t \sim 47$ ps OH elimination takes place.  Black, red, and
      green solid lines correspond to the C$_{\rm A}$H$_{\rm A}$,
      O$_{\rm A}$O$_{\rm B}$, and O$_{\rm B}$H$_{\rm A}$ distances,
      respectively.}
    \label{fig:tser}
\end{figure}

\noindent
{\bf A Typical Reactive Trajectory} An illustrative example for a
reactive trajectory from MS-ARMD simulations is reported in Figure
\ref{fig:tser}A. Here, the C$_{\rm A}$H$_{\rm A}$, O$_{\rm B}$H$_{\rm
  A}$, and O$_{\rm A}$O$_{\rm B}$ time series are shown (for labeling
see Figure \ref{fig:pes}A) which are directly relevant to the
reaction. Initially, the C$_{\rm A}$H$_{\rm A}$ separation (black
trace in Figure \ref{fig:tser}A) fluctuates around 1.12 \AA\/ which is
close to the equilibrium bond length, and the O$_{\rm B}$H$_{\rm A}$
separation is large (ranging from 2 to 4 \AA\/) as the system is in
its reactant state. Such large variations are elicited by the CH$_3$
rotation. At $t \sim 0.5$ ns the 1,4 hydrogen shift reaction occurs
which increases the C$_{\rm A}$H$_{\rm A}$ and O$_{\rm A}$O$_{\rm B}$
separations and shortens the O$_{\rm B}$H$_{\rm A}$ bond due to bond
formation. For the next $\sim 0.3$ ns the system is in its VHP state
before OH elimination takes place at $t \sim 0.79$ ns following
breaking of the O$_{\rm A}$O$_{\rm B}$ bond.\\

\noindent
A similar trajectory, run with ASE and the PhysNet representation, is
shown in Figure \ref{fig:tser}B. Again, CH$_3$ rotation is found. As
for the MS-ARMD simulations, for VHP (between $t \sim 28$ ps and $t
\sim 47$ ps) the position of the OH-group switches between {\it syn-}
and {\it anti-}conformers, respectively. These lead to long and short
C$_{\rm A}$--H$_{\rm A}$O$_{\rm B}$ separations (black trace).\\

\noindent
{\bf The Vibrationally Assisted Reaction} In laboratory-based
experiments\cite{fang:2016,lester:2016} the reaction path following
\textit{syn}-CH$_3$CHOO $\rightarrow$ VHP $\rightarrow$ CH$_2$CHO+OH
is initiated by excitation of the CH stretch vibration of the terminal
CH$_3$ group with energies ranging from $\sim 5600$ cm$^{-1}$ to $\sim
6000$ cm$^{-1}$. This corresponds to about 2 quanta in the methyl-CH
stretch mode. In the simulations, excitation of this mode was
accomplished by scaling the
velocities\cite{reyes.pccp.2014.msarmd,reyesbrickel.pccp.2016.msarmd}
along the CH-local mode. Equilibrium simulations for
\textit{syn}-CH$_3$CHOO were carried out at 300 K and 50 K to generate
the initial ensemble. It should, however, be noted that
``temperature'' as determined from the equivalence of kinetic energy
and $3/2N k_{\rm B}T$, as is usually done in MD
simulations,\cite{schmelzer:2010} should not be directly compared with
experimentally reported temperatures (e.g. 10 K rotational temperature
in Ref.\cite{fang:2016}).\\

\begin{figure}[h!]
   \centering \includegraphics[scale=0.49]{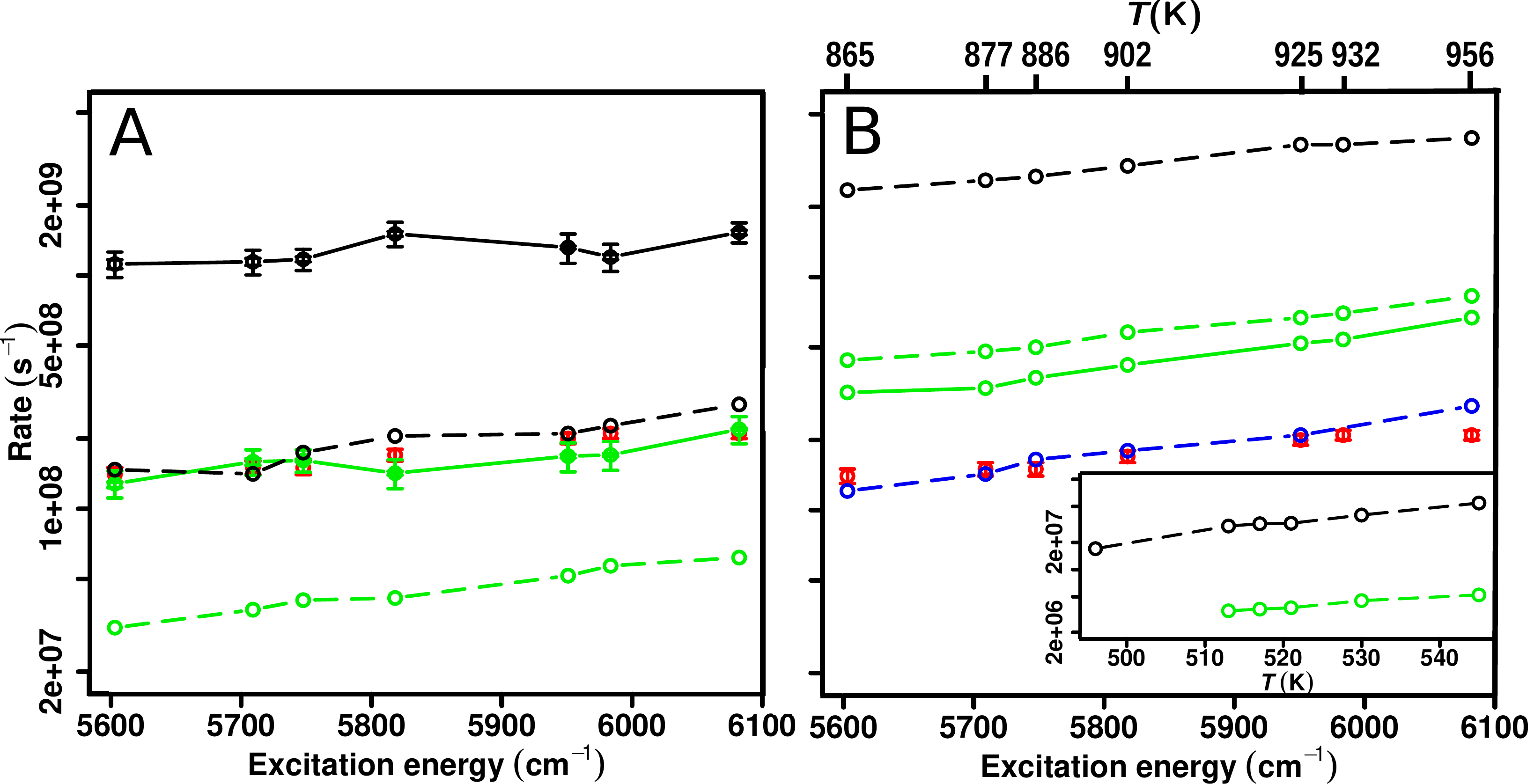}
  \caption{OH formation rates from vibEX and thermal
    simulations. Panel A: Rates from MS-ARMD simulations following
    excitation of the C$_{\rm A}$H$_{\rm A}$ stretch vibration (vibEX)
    at different excitation frequencies. The green and black lines are
    for $D_{\rm e}^{\rm OO} = 31.5$ kcal/mol (consistent with the
    present and earlier\cite{kurten:2012} MP2 calculations), and
    $D_{\rm e}^{\rm OO} = 23.5$ kcal/mol respectively. The red points
    with error bars are the experimental results. Solid and dashed
    lines are rates from fitting $N(t)/N_{\rm tot}$ to stretched or
    single exponentials, respectively, with separate error bars from
    bootstrapping and for the parameters. Panel B: Rates from thermal
    simulations using MS-ARMD (identical color code as in panel A),
    and from PhysNet (dashed blue line). The top $y-$axis gives the
    temperature $T$ as determined from $T = 2E_{\rm kin}/(3 N k_{\rm
      B})$. The inset reports results from simulations at somewhat
    lower temperatures.}
    \label{fig:rates}
\end{figure}

\noindent
Simulations were run using the same internal energies as those
reported from experiments, i.e. [5603, 5709, 5748, 5818, 5951, 5984,
  6082] cm$^{-1}$. For each energy $N_{\rm tot} = 10000$ individual
trajectories were run. The ensuing rates were determined from
following the number $N(t) = 1 - N_{\rm OH }$, i.e. counting those
that had not reached the OH-elimination product, see Table
S1 and S2 and fitting $N(t)$ to
either a single ($\sim \exp{(-kt)}$) or a stretched-exponential
($\exp{(-kt)^\gamma}$) dependence (see Figure
S3).\cite{MM.mbno:2016} The rates $k$ as a
function of excitation energy are reported in Figure \ref{fig:rates}A
together with those from experiment (red symbols).\cite{fang:2016}\\

\noindent
MS-ARMD simulations from an ensemble generated at 300 K were carried
out for two values of the O--O dissociation energy, see above. Rates
from simulations with $D_e^{\rm OO} = 31.5$ kcal/mol (green) and
$D_e^{\rm OO} = 23.5$ kcal/mol (black) are shown in Figure
\ref{fig:rates}A. Depending on the value of $D_e^{\rm OO}$ the rates
differ by a factor of $\sim 10$ but following a similar energy
dependence, consistent with that observed in the
experiments. Bootstrapping is used to determine statistical errors by
randomly sampling 8000 trajectories 30 times from all 10000
trajectories at each energy. The error bars due to parameter
uncertainties in the fitting of $\ln{N(t)/N(0)}$ vs. time are a factor
of 4 to 10 smaller than those from bootstrapping.\\

\noindent
To assess the sensitivity of the results to the initial preparation of
the system, vibEX dissociation simulations were also carried out by
sampling from an ensemble generated at 50 K using MS-ARMD and
$D_e^{\rm OO} = 23.5$ kcal/mol. Several 1000 simulations were run for
5 ns at different excitation energies. At 5603 cm$^{-1}$ 15 out of
2000, 5747 cm$^{-1}$ 23 out of 1000, 5818 cm$^{-1}$ 6 out of 400 show
VHP formation. Hence, reactivity is also found for considerably lower
(thermal) energies but the rates are slower than those started from
the ensemble generated at 300 K. However, a larger number of
trajectories would be required for better converged rates.\\

\noindent
{\bf The Thermal Reaction} In the atmosphere, vibrational excitation
is not the likely primary mechanism by which \textit{syn}-CH$_3$CHOO
is energized. Rather, the ozonolysis reaction is
expected\cite{kroll:2001} to lead to a ``warm'' or ``hot'' parent
molecule (\textit{syn}-CH$_3$CHOO) which subsequently decays following
the 1,4 hydrogen shift and O--O dissociation
reactions.\cite{donahue:2011,taatjes:2015,wang:2019} Hence,
``thermal'' simulations were also carried out by heating the reactant
to average temperatures between 865 K and 956 K, commensurate with
internal energies of $\sim 5500$ to 6000 cm$^{-1}$. Again, the
equivalence between ``temperature'' and ``kinetic energy'' is rather
qualitative. For this reason, additional simulations at somewhat lower
temperatures (500 K to 550 K) were also carried out and analyzed.\\

\noindent
The rates were determined by following $N(t)$, see Tables
S3, S4 and S5. For
these ``thermal'' simulations with $D_e^{\rm OO} = 31.5$ kcal/mol,
$N(t)$ is well represented by a single exponential decay for times $t
< 200$ ps whereas a stretched exponential is again required for longer
simulation times, see Figure S4. Such information
is of particular relevance in the context of {\it ab initio} MD
simulations which can usually only be carried out on the multiple
$\sim 10$ ps time scale for sufficiently high-level treatments of the
electronic structure.\cite{gerber:2014}\\

\noindent
Thermal rates from $\sim 9000$ independent simulations using the
MS-ARMD (solid green and black lines) and PhysNet (dashed blue line)
representations are reported in Figure \ref{fig:rates}B. MS-ARMD
simulations were run for the same two values of the O--O dissociation
energy as the vibEX simulations with rates ranging from $3\times10^8$
to $5\times10^9$ s$^{-1}$. The temperature dependence is moderate and
follows that found from experiments using vibrational excitation
although the magnitude of the computed rate is consistently higher by
a factor of 5 to 10. Thermal rates from simulations using PhysNet show
the same temperature dependence as the one observed in experiments and
from MS-ARMD simulations but with an amplitude that is closer to that
from experiment ($\sim 10^8$ s$^{-1}$). Exploring effects due to a
change in $D_{\rm e}^{\rm OO}$ is not easily accomplished using
PhysNet because the underlying data set used in the fitting would need
to be modified accordingly.\\

\noindent
MS-ARMD simulations were also carried out for ensembles generated at
lower temperatures, see inset Figure \ref{fig:rates}. As expected, the
rates decrease by about an order of magnitude but remain consistent
with experiment, exhibiting a comparable temperature dependence.\\

\begin{figure}
    \centering \includegraphics[scale=0.32]{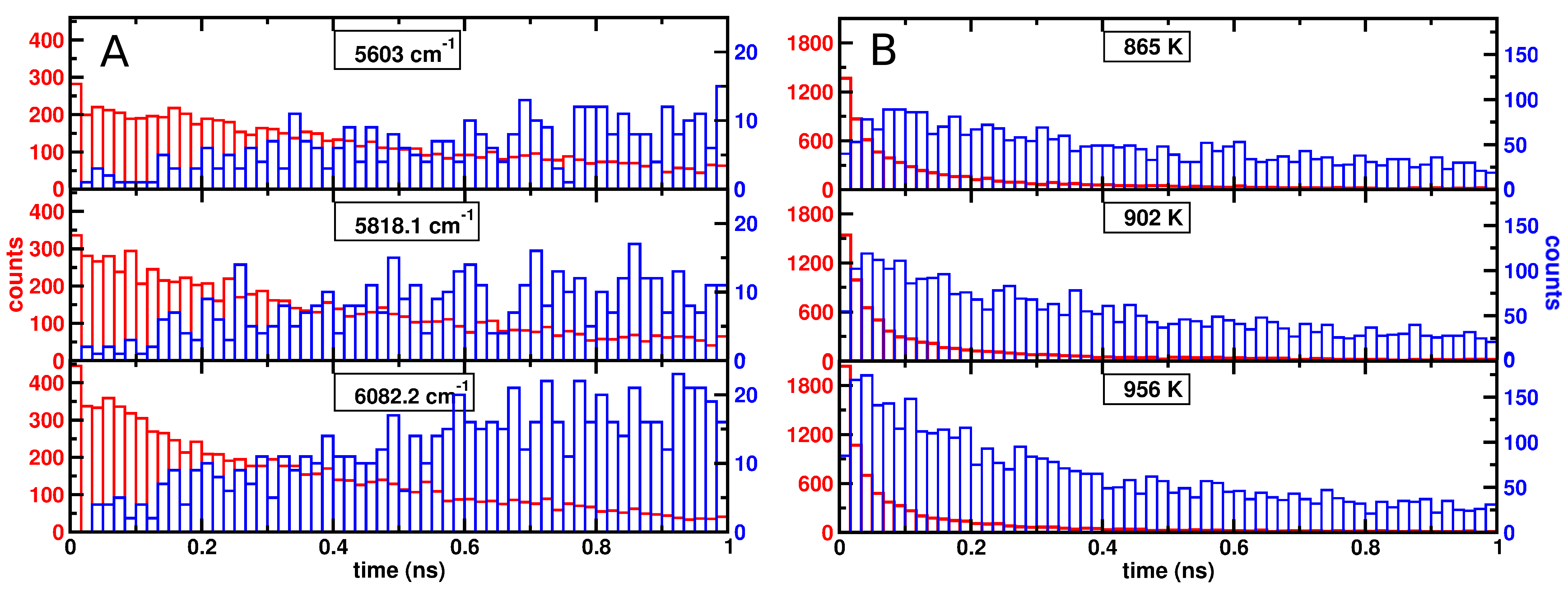}
    \caption{Distribution of reaction times for 1,4 H-shift
      [CH$_3$CHOO$\rightarrow$VHP, red] and OH-formation
      [CH$_3$CHOO$\rightarrow$CH$_2$COH+OH, blue] from vibEX (Panel A)
      and thermal (Panel B) simulations at different energies and
      temperatures using MS-ARMD with $D_{\rm e}^{\rm OO} = 31.5$
      kcal/mol. Note that scales for the 1,4 H-shift distribution
      times are along the left hand $y-$axis and those for the total
      reaction time are along the right hand $y-$axis and that the
      magnitudes differ between panels A and B.}
    \label{fig:time}
\end{figure}

\noindent
{\bf Analysis of the Reactive Trajectories} With a statistically
significant number of reactive trajectories ($\sim 10^4$) it is also
possible to carry out additional analyses. Distributions of H-transfer
and OH-formation times from vibEX and thermal MS-ARMD simulations are
reported in Figure \ref{fig:time}. Reactions initiated from
vibrational excitation have a comparatively flat reaction time
distribution up to $\sim 0.2$ ns for the 1,4 H-shift reaction, red
histogram in Figure \ref{fig:time}A. The reaction time is defined by
the time interval between the point of vibrational excitation and VHP
formation for which a geometrical criterion was used (O-H separation
$< 1.6$ \AA\/ and O-O separation $< 2.8$ \AA\/). For longer simulation
times ($> 0.2$ ns) the reaction probability decays towards zero. Decay
to CH$_2$COH+OH in the vibEX simulations is delayed and the number of
trajectories reaching the final state compared with those that
complete the first step (1,4 H-shift) depends on time. For lower
excitation energy (top panel in Figure \ref{fig:time}A) appreciable OH
elimination only starts after $\sim 0.2$ ns and does not reach more
than 10 \% of the amount of VHP formed on the 1 ns time scale. For
higher excitation energies, the onset of OH elimination shifts to
shorter times and the amount of OH formed increases in proportion. The
fraction of VHP intermediates formed ranges from 74 \% to 87 \%
compared with a fraction of 3 \% and 7 \% for the amount of OH formed
from vibEX simulations (5603 cm$^{-1}$ to 6082.2
cm$^{-1}$). Therefore, an appreciable amount of VHP accumulates on the
1 ns time scale and reacts to the OH-elimination product on longer
time scales.\\

\noindent
The situation changes considerably for the thermal reaction, see
Figure \ref{fig:time}B. Here, the reaction time distributions for the
1,4 H-shift reaction approach an exponential decay with simulation
time for all temperatures considered. Similarly, the amount of OH
formed follows this. With increasing temperature (internal kinetic
energy) the amount of VHP formed increases by about 50 \% between 865
K and 956 K and the amount of OH formed by almost a factor of two. The
fraction of VHP intermediates formed ranges from 73 \% to 76 \%
compared with a fraction of 28 \% and 38 \% for the amount of OH
formed.\\

\noindent
The results from Figure \ref{fig:time} indicate that the VHP
intermediate has a broad distribution of lifetimes, see Figure
S6. As the probability to form VHP in the vibEX
simulations has not reached zero within 1 ns (red distribution Figure
\ref{fig:time}A), VHP can also be generated on longer time scales from
trajectories that have undergone more extensive internal vibrational
energy redistribution (IVR). For the thermal trajectories the
probability to form VHP has decayed to very low levels within 1 ns
(Figure \ref{fig:time}B). Thus, the VHP lifetime distributions in
Figure S6 are expected to be
close-to-converged. Importantly, vibrational excitation leads to about
one order of magnitude less OH product than thermal preparation of the
system at a similar internal energy.\\

\begin{figure}
    \centering \includegraphics[scale=0.35]{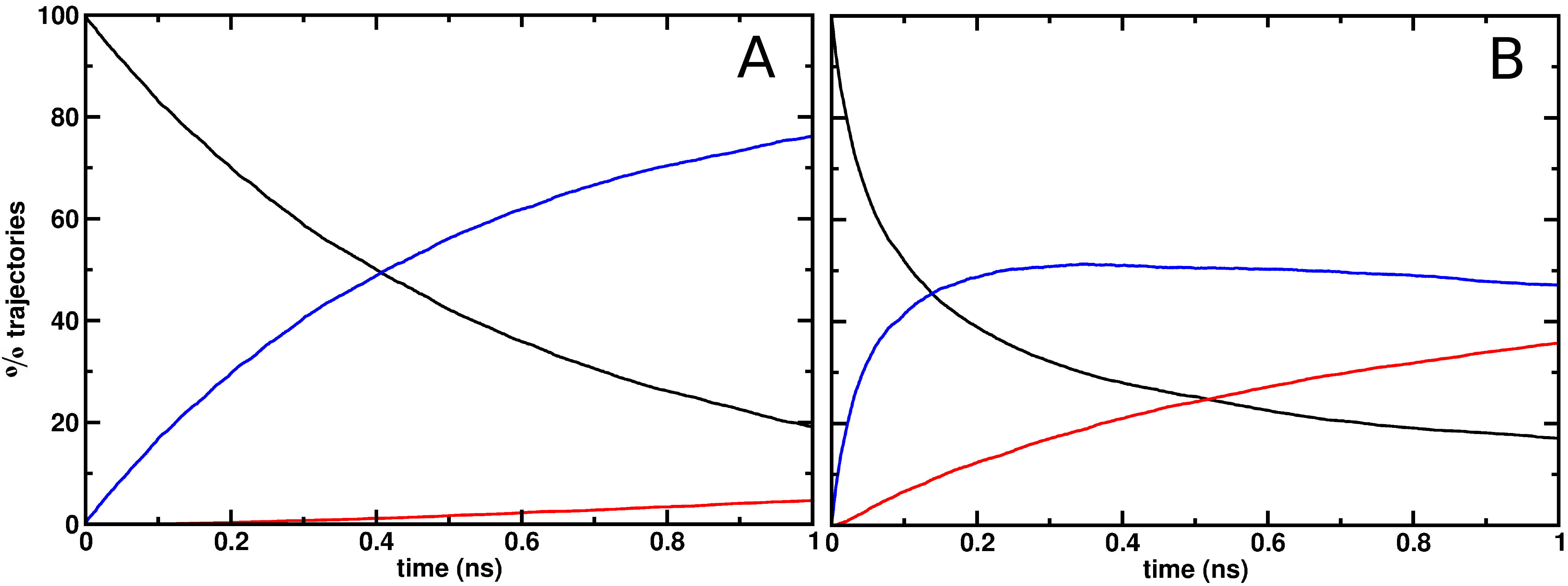}
    \caption{Change in concentration (in \%) of Criegee Intermediate,
      VHP and CH$_2$COH+OH as a function of time using MS-ARMD and
      $D_e^{\rm OO} = 31.5$ kcal/mol. Panel A: from vibEX simulations
      at 5818.1 cm$^{-1}$. Panel B: from thermal simulations at 902 K
      for $N_{\rm Criegee}$ (black), $N_{\rm VHP}$ (blue), and $N_{\rm
        OH}$ (red), respectively. For vibEX the rate limiting step is
      the OH-elimination step whereas for thermal excitation none of
      the two steps is a clear reaction bottleneck. With comparable
      energy content in the reactant, thermal simulations yield close
      to one order of magnitude more OH-product than vibrational
      excitation of the methyl-CH-stretch.}
    \label{fig:counts}
\end{figure}

\noindent
Formation of the VHP intermediate follows different time traces
depending on whether energy to {\it syn-}CH$_3$CHOO was provided by
methyl-CH vibrational excitation or thermal preparation. Figure
\ref{fig:counts}A shows that VHP concentration from vibEX
monotonically increases (blue trace) on the 1 ns time scale and
reaches close to 80 \% but only for 5 \% OH elimination has
occurred. Thermal excitation - representative of the initial state of
the reactant after ozonolysis of trans-2-butene - leads to $\sim$ 50\%
population of VHP within 0.3 ns. Over the same time close to 20 \% of
the trajectories already show OH-elimination. Up to 1 ns more than 30
\% of {\it syn-}CH$_3$CHOO have completed OH-elimination. Hence, OH
production on the 1 ns time scale is about one order of magnitude
larger with thermal compared to vibEX preparation of the reactant.\\

\begin{figure}[h!]
   \centering \includegraphics[scale=0.48]{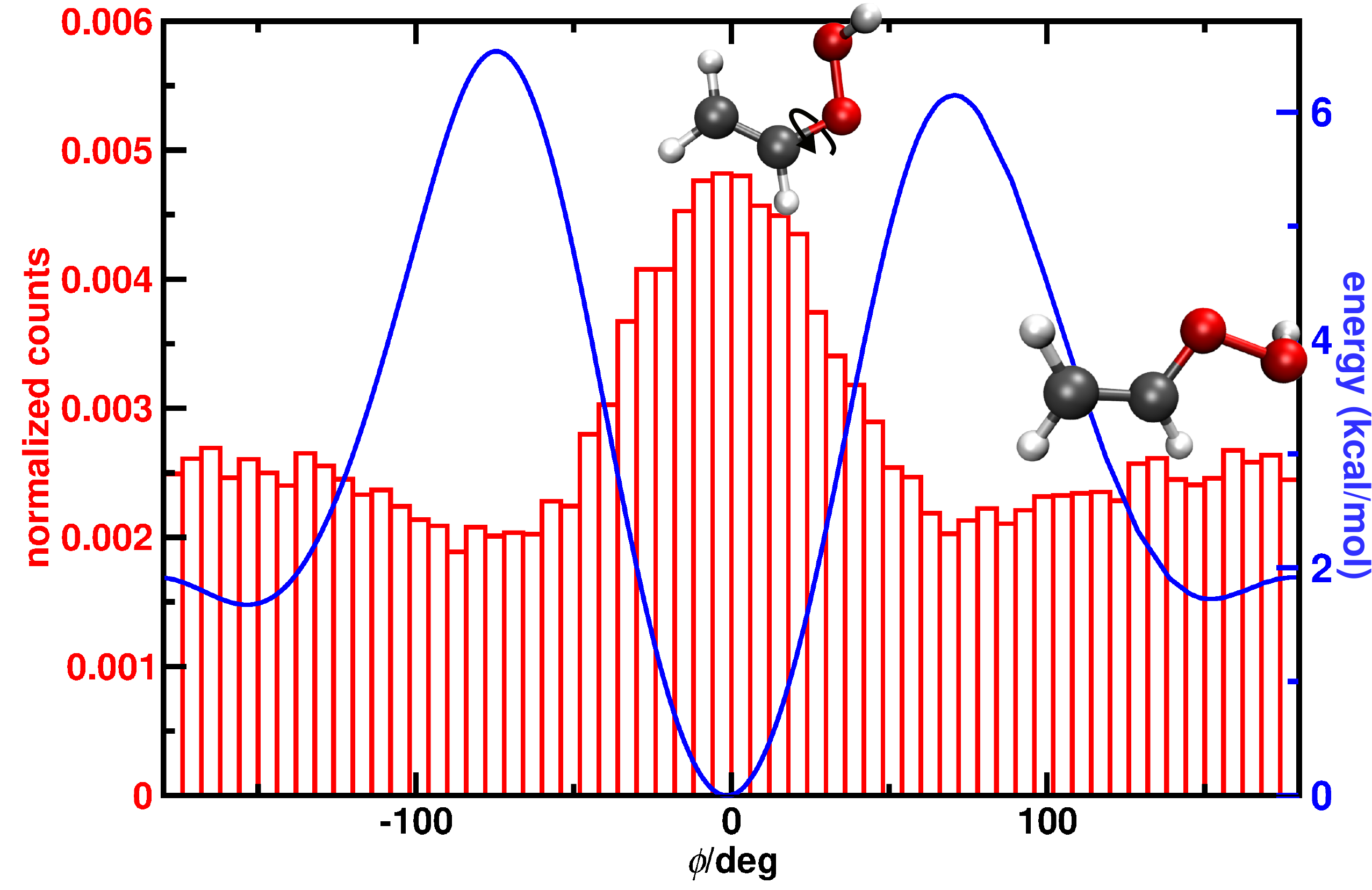}
  \caption{The CCOO dihedral angle distribution $P(\phi)$ for
    CH$_2$CHOOH (VHP) from thermal simulations at 902K using ASE. The
    structure with $\phi = 0^\circ$ corresponds to the -OH group in
    its \textit{syn} conformation whereas $\phi = \pm 180^\circ$
    corresponds to the \textit{anti} conformer. The barrier between
    $0^\circ$ and $180^\circ$ is 6.1 kcal/mol and from $0^\circ$ to
    $-180^\circ$ is 6.5 kcal/mol at the MP2/aug-cc-pVTZ level of
    theory. The histogram of population for the VHP conformation from
    the simulations is consistent with the PES as \textit{syn}- form
    is more stable than \textit{anti}- form.}
    \label{fig:dihedral}
\end{figure}

\noindent
The extended lifetimes of VHP prior to OH elimination can also be
rationalized from considering the CCOO dihedral angle $\phi$, see
Figure \ref{fig:dihedral}. Coming from CH$_3$CHOO the VHP intermediate
is formed with a considerable amount of internal energy which is
sufficient to overcome the syn/anti barrier height of $\sim 6$
kcal/mol, see also Figure \ref{sifig:dihedral}. The population
distribution $P(\phi)$ for VHP from simulations with PhysNet
simulations at 902 K sample all values of $\phi$, consistent with the
topology of the potential energy curve (blue trace in Figure
\ref{fig:dihedral}). Analysis of the simulations reveals that
OH-elimination occurs from both, {\it syn-} and {\it anti-}VHP. This
is consistent with the energetic preference for {\it syn-}VHP but at
variance with findings from statistical modeling which reported that
all of the VHP decomposes via the {\it anti-}pathway even if the {\it
  syn} form is thermodynamically more stable.\cite{kuwata:2018}\\

\noindent
Finally, bond length distributions (Figures S8 and
S9) show that there is little difference between vibEX
and thermal simulations (blue compared with red traces). Thermal
distributions from MS-ARMD simulations using only the reactant force
field at 300 K (black) are considerably more peaked around the
equilibrium values compared with those at 760 K (red). If the 1,4
H-shift reaction is possible (by running the MS-ARMD simulations with
the reactive force field), a few bond length distributions change
appreciably (green traces in Figure S8). Excitation of
the C$_{\rm A}$H$_{\rm A}$ bond leads to a considerably flatter, more
extended distribution, similarly to the C$_{\rm B}$H$_{\rm D}$
distribution. Conversely, the C$_{\rm A}$C$_{\rm B}$ distribution
function slightly narrows with an increased peak height at the minimum
energy geometry. This suggests that compression of the C-C bond
facilitates the 1,4 H-shift reaction.\\

\section*{Discussion and Conclusion}
The present work reports on the OH-elimination reaction dynamics of
\textit{syn}-CH$_3$CHOO at thermal and vibrationally induced initial
conditions. The thermal process is relevant under conditions that
follow the ozonolysis reaction of trans-2-butene which generates a
``warm'' or ``hot'' reactant. Vibrational excitation of the
methyl-CH-stretch is used in gas-phase laboratory experiments to
initiate the reaction. All simulations carried out in the present work
find ready 1,4-H-transfer to form the VHP intermediate which then
partially reacts to products CH$_2$COH+OH. One of the important
``unknowns'' remains the dissociation energy $D_e^{\rm OO}$ which
would require large-scale multi reference configuration interaction
calculations. Of particular note is the finding that tunneling is not
required for the entire pathway to obtain rates consistent with
experiment.\\

\noindent
Because the reaction involves a 1,4 H-shift it is quite likely that
tunneling will contribute to the rate. However, it is not the
determining factor for reactivity. This differs from earlier efforts
based on RRKM theory that reported appreciable rates only when
tunneling was included.\cite{fang:2016} To further probe this,
additional MS-ARMD simulations with $D_e^{\rm OO} = 31.5$ kcal/mol
were carried out that with excitation below the barrier. Starting from
samples at 300 K with 4000 cm$^{-1}$ of excess energy, OH-elimination
occurs on the 2 ns time scale. With the lower, probably preferred,
value for $D_e^{\rm OO}$ the reaction is expected to proceed even more
readily.\\

\noindent
The PESs used here allow to run a statistically significant number of
reactive trajectories on the nanosecond time scale with qualities
approaching the MP2 level of theory at the cost of an empirical force
field. The two representations have their particular advantages and
shortcomings. The MS-ARMD PES has an overall accuracy of $\sim 1$
kcal/mol which is certainly sufficient for qualitative and
semi-quantitative studies. Such a parametrization allows exploration
of parameter space as illustrated by the variation of the well depth
$D_e^{\rm OO}$. On the other hand, the PhysNet representation is
highly accurate with respect to the reference points. Simulations with
this PES are about two orders of magnitude slower which limits broad
exploration of initial conditions. One example concerns vibEX
simulations with PhysNet. These needed to be carried out with
excitation energies ranging from 6500 cm$^{-1}$ to 9000 cm$^{-1}$ in
order to observe reactive trajectories due to the O--O dissociation
energy of $D_e^{\rm OO} = 35.7$ kcal/mol. With 9000 cm$^{-1}$ excess
energy 1 out of 10 trajectories show OH-elimination on the 1 ns time
scale whereas with 8000 cm$^{-1}$ 1 out of 5 trajectories progressed
to product on the 5 ns time scale. Excitation with 6500 cm$^{-1}$ does
not lead to OH-elimination on the 25 ns time scale. Exploration of the
influence of $D_e^{\rm OO}$ within PhysNet is not easily possible
without dedicated modification of the underlying data set and training
a new NN.\\

\noindent
The vibEX simulations indicate that the 1,4 H-shift reaction yields
$\sim 80$ \% VHP on the 1 ns time but only 5 \% react further to the
product, see Figure \ref{fig:counts}A. Therefore, the second step is a
bottleneck for OH generation following vibrational excitation. This is
at variance with earlier reports that favour prompt OH-loss and find
that the 1,4 H-shift reaction is rate limiting (based on RRKM studies
and dynamics initialized at the transition state between
\textit{syn}-CH$_3$CHOO and VHP)\cite{fang:2016,lester:2016} but
consistent with experimental evidence for significant collisional
stabilization of VHP prior to OH
formation.\cite{drozd:2011,donahue:2011} One possible explanation for
the results found here is rapid IVR after the 1,4 H-shift
reaction. This can be seen, e.g., in the high excitation of the
CCOO-dihedral motion after formation of VHP (Figure
\ref{sifig:dihedral}). Similar observations were made for the
isomerization of acetaldehyde (AA) to vinylalcohol
(VA).\cite{MM.atmos:2020} Excitation of AA with an actinic photon
($\sim 94$ kcal/mol) is not sufficient to trigger isomerization to VA
on the 500 ns time scale although the AA$\rightarrow$VA barrier height
is only 68 kcal/mol. Conversely, the thermal simulations which are
representative of initial conditions following ozonolysis of
trans-2-butene find that on the 1 ns time scale similar amounts of VHP
and OH-elimination products are formed with only $\sim 20$ \% of
reactant remaining, see Figure \ref{fig:counts}B. Hence, on this time
scale no clear bottleneck can be identified. How much VHP finally
reacts to form OH product also depends on the collisional quenching
time which is between 1 ns and 10
ns.\cite{lester:2017,reyes.pccp.2014.msarmd} Hence, if VHP does not
form OH product on that time scale it is more likely to loose energy
in collisions with the environment which limits OH production from
this pathway.\\

\noindent
OH-elimination from \textit{syn}-CH$_3$CHOO following vibrational
excitation or thermal preparation yields rates consistent with
experiments using full dimensional MS-ARMD and NN-based PESs. The
classical MD simulations do not include tunneling effects which are
expected to further speed up the first step. Following vibrational
excitation of the CI, VHP is found to accumulate. This is different
for thermal preparation of {\it syn-}CH$_3$CHOO. Overall, the present
work provides molecular-level detail for an important reaction in
atmospheric chemistry. The approaches used here are generic and
expected to be applicable to a range of other reactions.\\

\section*{Data Availability Statement}
The PhysNet codes are available at
\url{https://github.com/MMunibas/PhysNet}, and the datasets containing
the reference data can be obtained from from github
\url{https://github.com/MMunibas/Criegee.git}.\\

\section*{Acknowledgments}
This work was supported by the Swiss National Science Foundation
through grants 200021-117810, 200020-188724 and the NCCR MUST, and the
University of Basel.

\section*{Supporting information}
The supplementary material contains the methods, tables with the
number of reactive trajectories, and figures for rate calculations,
lifetime statistics, and coordinate distributions functions.

\begin{figure}[h]
\centering \includegraphics[width=0.8\textwidth]{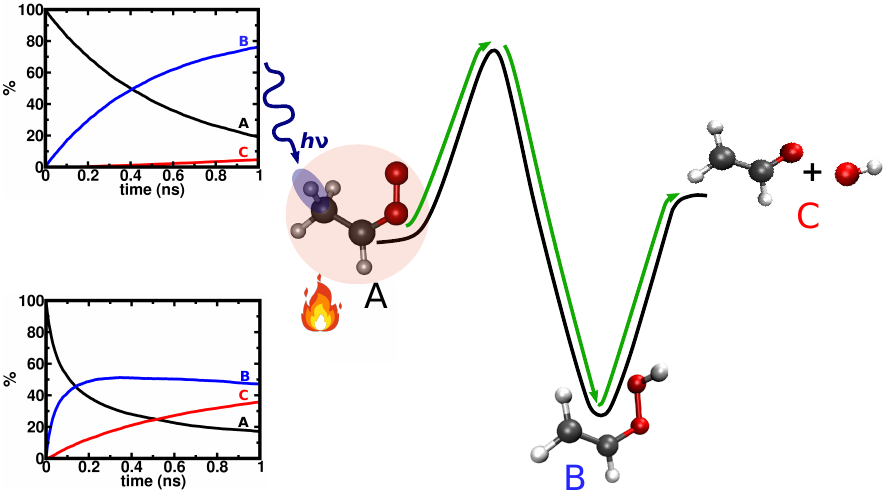}
\caption{Table of Contents graphics: Depending on the initial
  preparation the decomposition dynamics of {\it syn-}CH$_3$CHOO
  differs.}
\label{fig:toc}
\end{figure}

\bibliography{refs}
%\printbibliography
\end{document}

\section{Computational Methods}
This section presents the different computational methods
employed. Two different representations for the intermolecular
interactions are used. One is a multi-state reactive force field
(MS-ARMD), and the second one is a machine learning-based
representation using PhysNet. Finally, the molecular dynamics (MD)
simulations are also described.

\subsection{Reactive Force Fields}
MS-ARMD is a computationally efficient, energy-conserving surface
crossing algorithm to investigate chemical reactions based on
empirical force fields.\cite{MM.armd:2014} It uses parameterized force
fields for the reactant and product state and GAPOs
(GAussian×POlynomials) to describe the surface crossing region,
i.e. the region around the transition state (TS). The parametrized
force fields for the reactant and product complexes are iteratively
fit to reference data from electronic structure calculations.\\

\noindent
In the present work, two representations of the full-dimensional,
reactive PES were developed based on two quantum chemical
treatments. They included MS-ARMD and PhysNet for the representation
and the MP2/6-311++G(2d,2p) and MP2/aug-cc-pVTZ
levels of theory for the reference calculations which were carried out
using Gaussian09\cite{gaussian0920091} and
MOLPRO\cite{molpro:2020}. This allows validation of the
representations and direct assessment of the level of theory.\\

\noindent
The initial parameters for the reactant (\textit{syn}-CH$_3$CHOO,
methyl-substituted Criegee intermediate) and product (CH$_2$CHOOH,
vinyl hydroperoxide) were taken from
SwissParam.\cite{zoete2011swissparam} First, representative structures
for CH$_3$CHOO (1612) and CH$_2$CHOOH (1650) were sampled from 500 ps
MD simulations at 300 K. For these structures, energies were
determined at the MP2/6-311++G(2d,2p) level of theory. These were fit
separately to parametrized force fields for the reactant and the
product using a downhill simplex\cite{nelder1965simplex}
algorithm. Parametrization starts with a set of 100 structures for
CH$_3$CHOO and CH$_2$CHOOH and first fitting for the two states were
carried out which was then followed by further MD simulations using
this improved set of parameters from which another 200 structures was
extracted and included in the fit. Several rounds of refinements were
done until the root mean squared deviation for the final set between
the target (\textit{ab initio}) and the fitted energies for CH$_3$CHOO
and CH$_2$CHOOH reached 1.1 kcal/mol and 1.2 kcal/mol,
respectively. The bonds involved in bond breaking and bond formation
(the three C-H, the O-O, and the O-H bonds) were described by Morse
potentials and the charges from \textit{ab initio} calculations at the
MP2/6-311++G(2d,2p) level of theory. Generalized Lennard-Jones
potential is included and fitted, to represent the van der Waals
interaction between atoms(between H$_{\rm A}$ and O$_{\rm B}$ in
reactant FF and between H$_{\rm A}$ and C$_ {\rm A}$ in product FF) in
the reactive region.\\

\noindent
The reactant and product force fields are connected by using
GAPO-functions to yield a continuous connection along the reaction
path. For this, the intrinsic reaction coordinate (IRC) between
reactant and product was also determined and included in the
fitting. To parametrize the adiabatic barrier, genetic algorithm was
used to parametrize the GAussian×POlynomial (GAPO)
\cite{MM.armd:2014} functions and to reproduce the energies
along the reaction path. The TS barrier for \textit{syn}-CH$_3$CHOO is
16.02 kcal/mol.\\

\subsection{Neural Network}
As an alternative to MS-ARMD a NN-based reactive force field was
trained based on the PhysNet architecture,\cite{unke2019physnet}
 which is designed to construct PES after learning
molecular properties like energy, forces, charges and dipole moments
from ab initio reference data. Here, MP2/aug-cc-pVTZ level of
theory\cite{moller1934note} is used for generation of ab initio
reference data and the energies, forces and dipole moments are
calculated using MOLPRO software package.\\

\begin{figure}[!ht]
    \centering
    \includegraphics[scale=0.45]{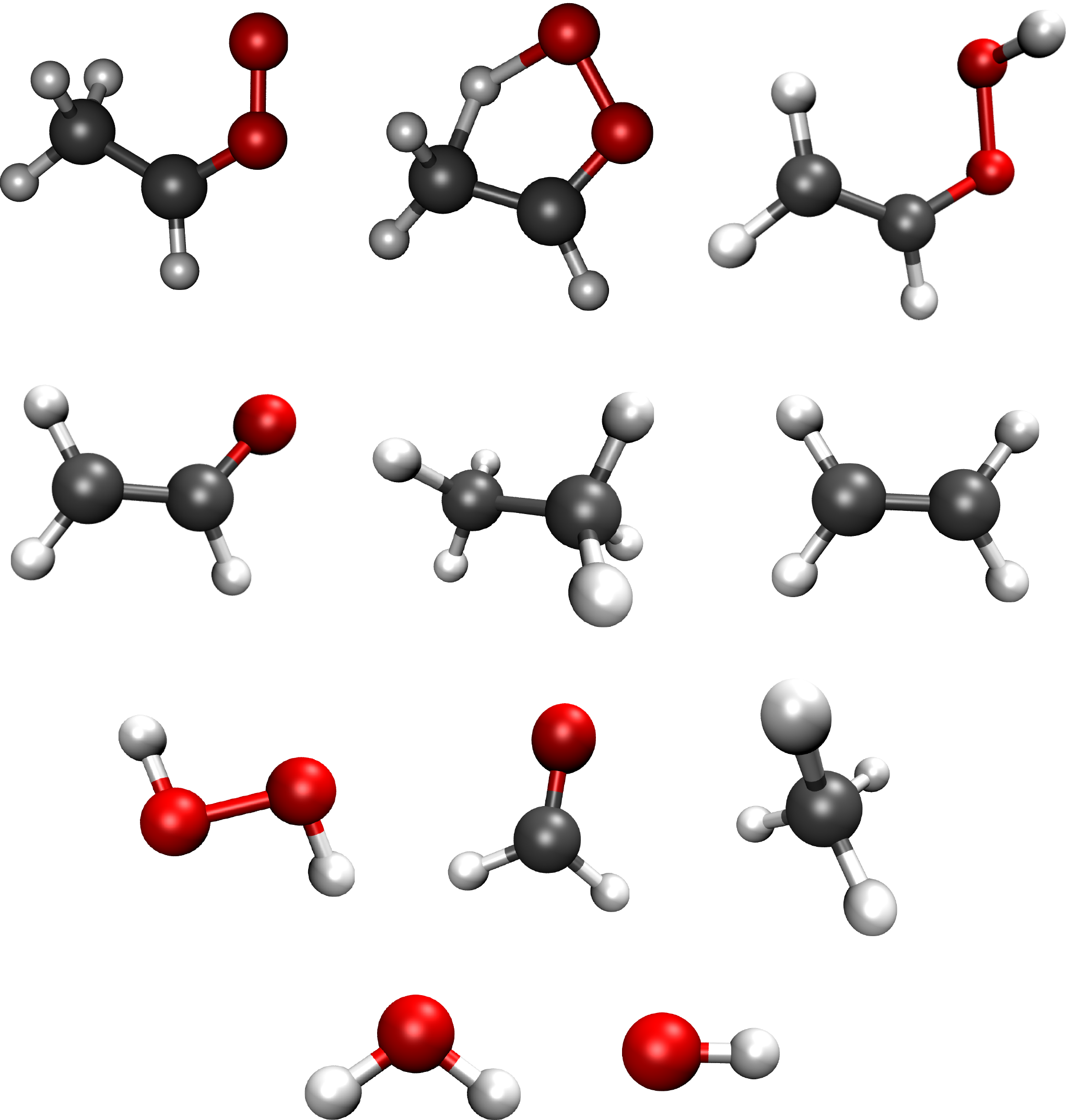}
    \caption{Different structures and amons\cite{oavl:2020amons} used
      for training the neural network representation. The first row
      contains the reactant, transition, product structures and
      subsequent rows contain their fragmented structures.}
    \label{fig:amons}
\end{figure}

\noindent
Following the ``amons" approach\cite{oavl:2020amons}, a set of
molecules (Figure \ref{fig:amons}) are included covering a range of
fragmentation reactant, products, stable intermediates and van der
Waals complexes in the dataset. To obtain a broad range of molecular
geometries, MD simulations were started from optimized geometries
which are propagated using Langevin dynamics at 1000 K with a time
step of 0.1 fs. Then, the data set is extended based on adaptive
sampling and normal mode sampling.\cite{behler2014representing} The
final dataset used for training contains 105403 structures, which was
then split into training (84322), validation (10540) and test set
(10541). The TS barrier for \textit{syn}-CH$_3$CHOO is 14.9
kcal/mol.\\

\subsection{Molecular Dynamics Simulations}
The MD simulations for MS-ARMD were carried out with a suitably
modified version of CHARMM.\cite{charmm:2009} Simulations based on
PhysNet were run with the Atom Simulation Environment
(ASE)\cite{larsen2017atomic}.\\

\noindent
For the thermal simulations based on MS-ARMD the geometry optimized
structure of CH$_3$CHOO was heated to the desired temperature and
equilibrated for 50 ps with a time step of 0.1 fs, followed by the 100
ps / 1 ns of free dynamics. The simulations with ASE were initialized
from the optimized structures. Then, momenta were assigned from a
Maxwell-Boltzmann distribution at the desired temperature which is
then followed by Langevin dynamics for 100 ps with $\Delta t = 0.1$
fs.\\

\noindent
The vibrationally excited (vibEX) simulations also started from a
geometry optimized structure of CH$_3$CHOO. Then, the system is heated
to 300 K and equilibrated for 50 ps with a time step of $\Delta t =
0.1$ fs, followed by 1 ns of free dynamics. From this simulation,
coordinates and velocities were saved regularly to obtain 70000
initial conditions for each of the excitation energies. Then a
non-equilibrium state is prepared by scaling the instantaneous
velocity vector along the OH mode. 10000 independent trajectories for
each excitation energies were run with a simulation time of 1 ns for
$D_e^{\rm OO} = 31.5$ kcal/mol and 100 ps for $D_e^{\rm OO} = 23.5$
kcal/mol. For the vibEX simulations with ASE, the momenta were
assigned from a Maxwell-Boltzmann distribution at 300 K to the
geometry optimized structure of CH$_3$CHOO, followed by 50 ps of free
dynamics with $\Delta t$= 0.1 fs. Then a non-equilibrium state is
prepared by scaling the instantaneous velocity vector along the CH
mode. 10000 independent trajectories were run for each excitation
energy for 100 ps with $\Delta t= 0.1$ fs.\\

\noindent
In addition to the explicit MD simulations, the minimum energy path
was also determined.\cite{MM.mdp:2019} This path connecting reactant
and product geometry, passing through the exact transition state with
zero excess energy was calculated by assigning momenta along the
normal mode vector. Then, the MD simulation from transition state to
both reactant and product channel using the PhysNet PES.\\

\begin{figure}[!ht]
    \centering
    \includegraphics[scale=0.36]{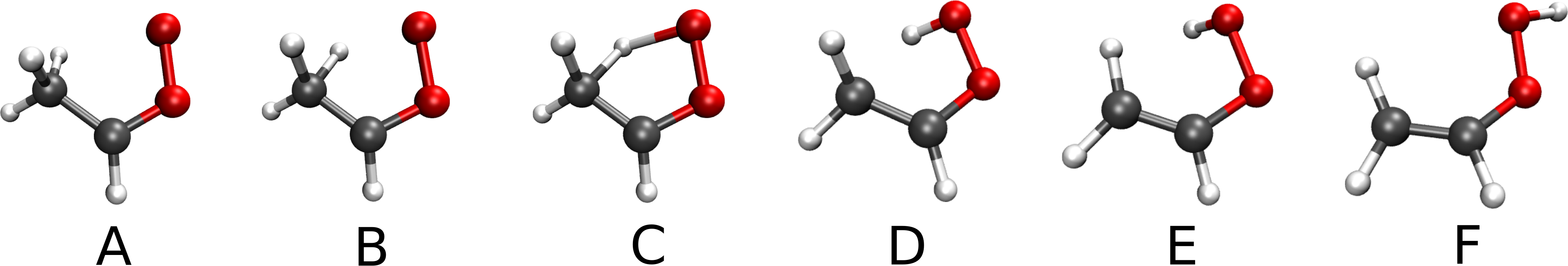}
 \caption{Minimum Dynamic Path\cite{MM.mdp:2019} on the PhysNet PES
   for the 1,4 Hydrogen shift from carbon to oxygen in Criegee
   intermediates. The reaction proceeds through a five membered
   transition state for which $r_{\rm CH}=1.30$ \AA\/ and $r_{\rm OH}
   = 1.41$ \AA\/. From A$\rightarrow$ F the C-C distance change is
   reported in Figure \ref{sifig:mdp-cc}. The C-O distance changes
   from 1.27 \AA\/ to 1.37 \AA\/ and the O-O distance changes from
   1.32 \AA\/ to 1.45 \AA\/.}
\label{sifig:mdp}
\end{figure}

\begin{figure}[!ht]
    \centering \includegraphics[scale=0.45]{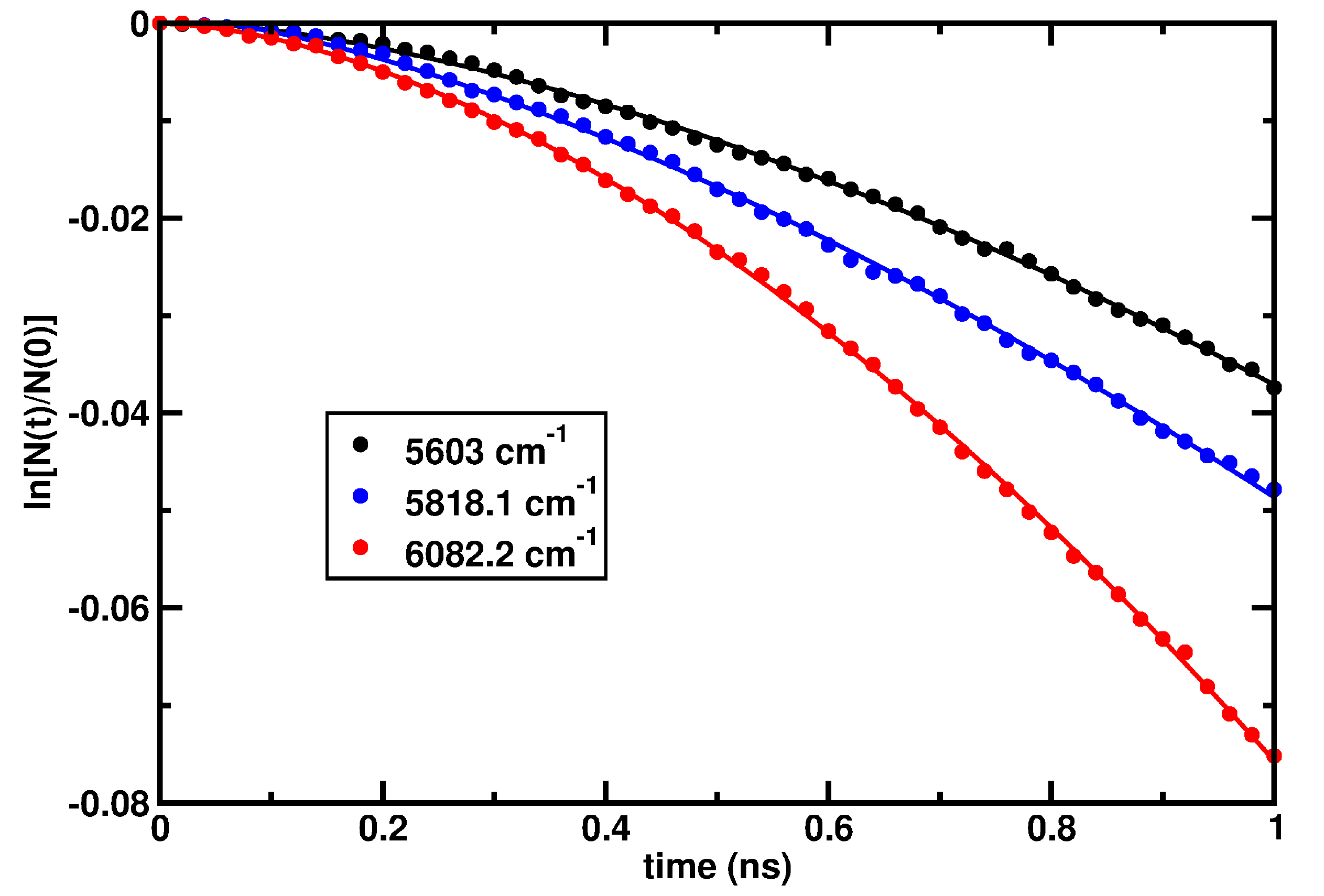}
    \caption{Fit of ln[$N(t)/N(0)$] for $D_e = 31.5$ kcal/mol from
      vibEX simulations at energies of 5603 cm$^{-1}$, 5818.1
      cm$^{-1}$ and 6082.2 cm$^{-1}$ to a stretched exponential
      decay.}
    \label{sifig:stretchfit}
\end{figure}

\begin{figure}[h!]
    \centering \includegraphics[scale=0.45]{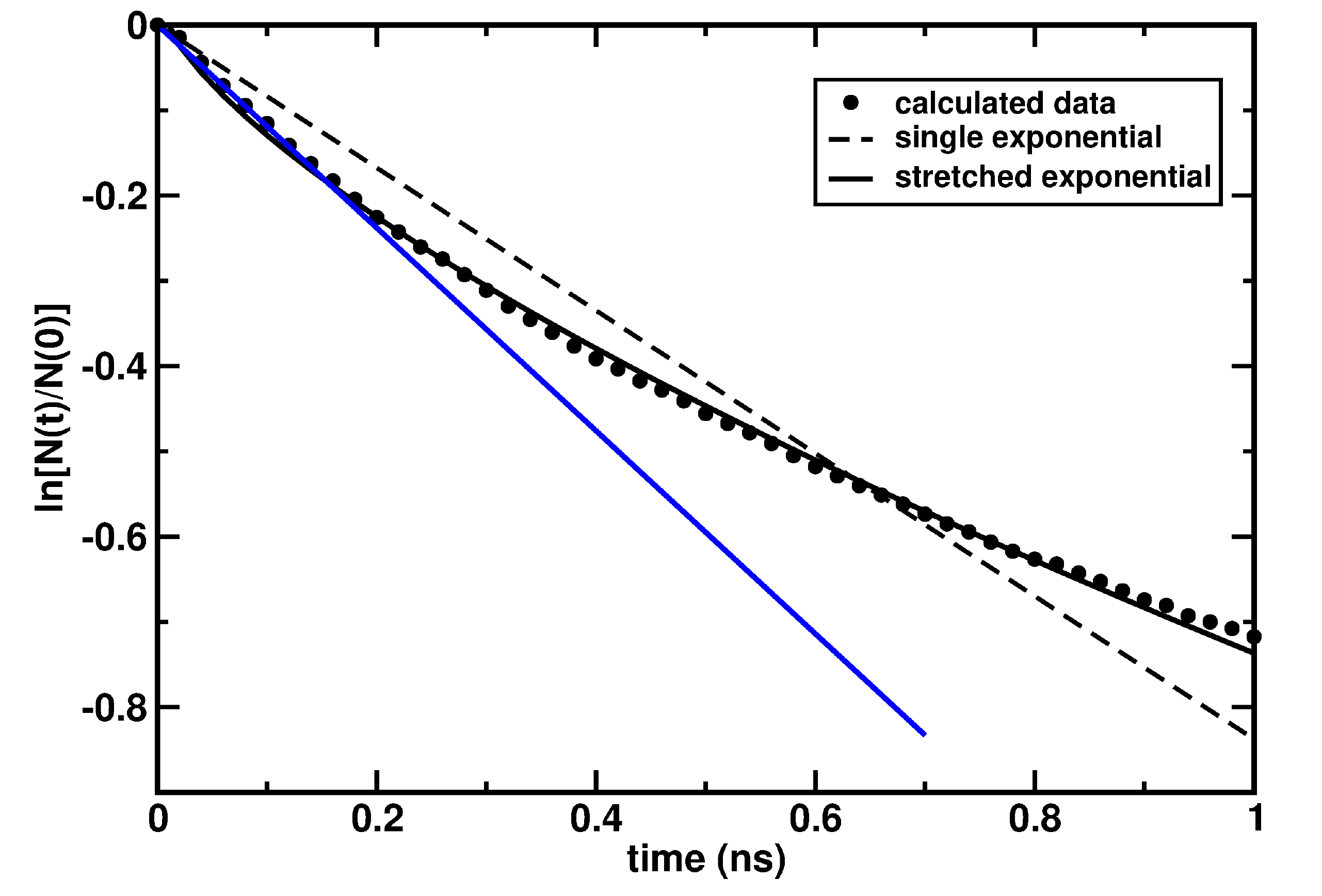}
    \caption{Fit of ln[$N(t)/N(0)$] for $D_e = 31.5$ kcal/mol from
      thermal simulations at 956 K where blue
      solid line is to show that till 200 ps the data follows a linear
      fit.}
    \label{sifig:k.thermal}
\end{figure}

\begin{figure}[h!]
    \centering
    \includegraphics[scale=0.45]{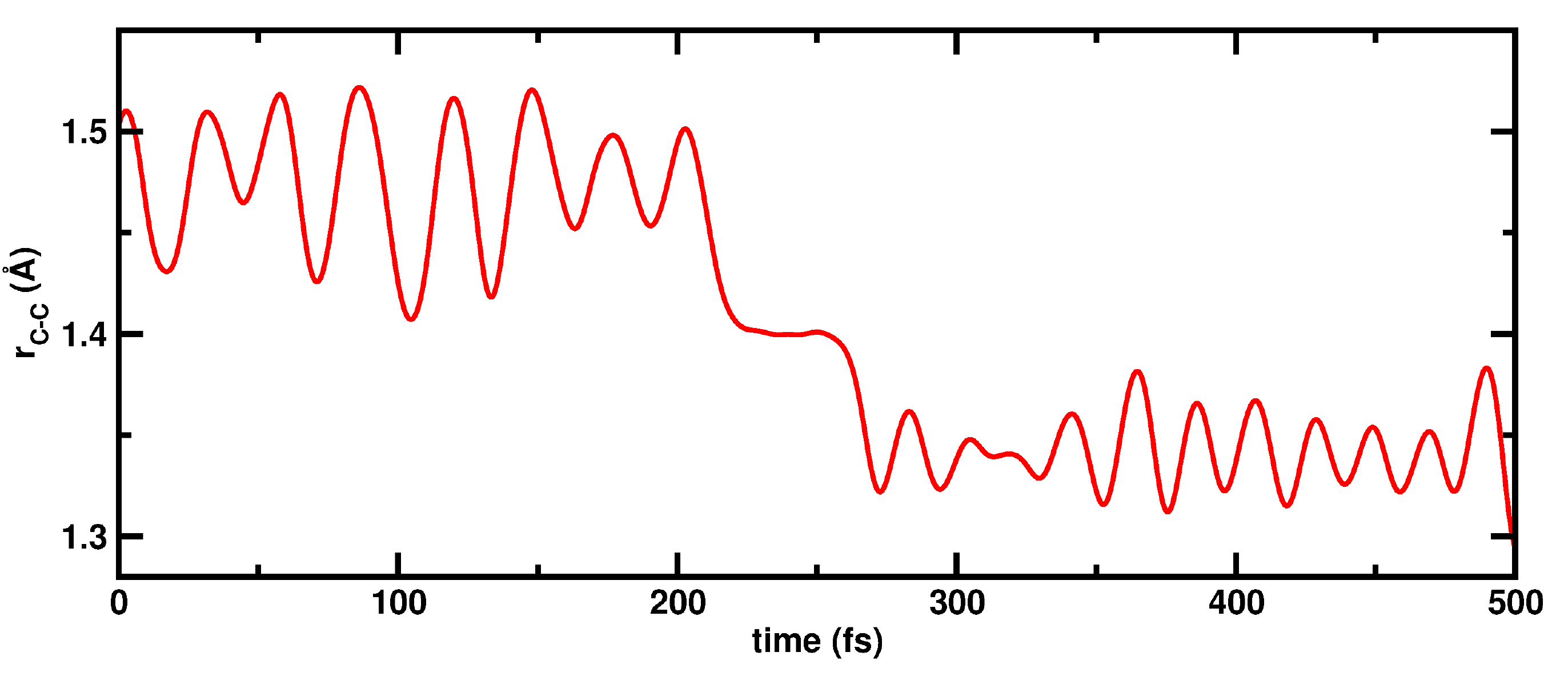}
    \caption{C-C distance along the minimum dynamic path as a function
      of time from CH$_3$CHOO to CH$_2$CHOOH.}
    \label{sifig:mdp-cc}
\end{figure}

\begin{figure}[h!]
    \centering \includegraphics[scale=0.32]{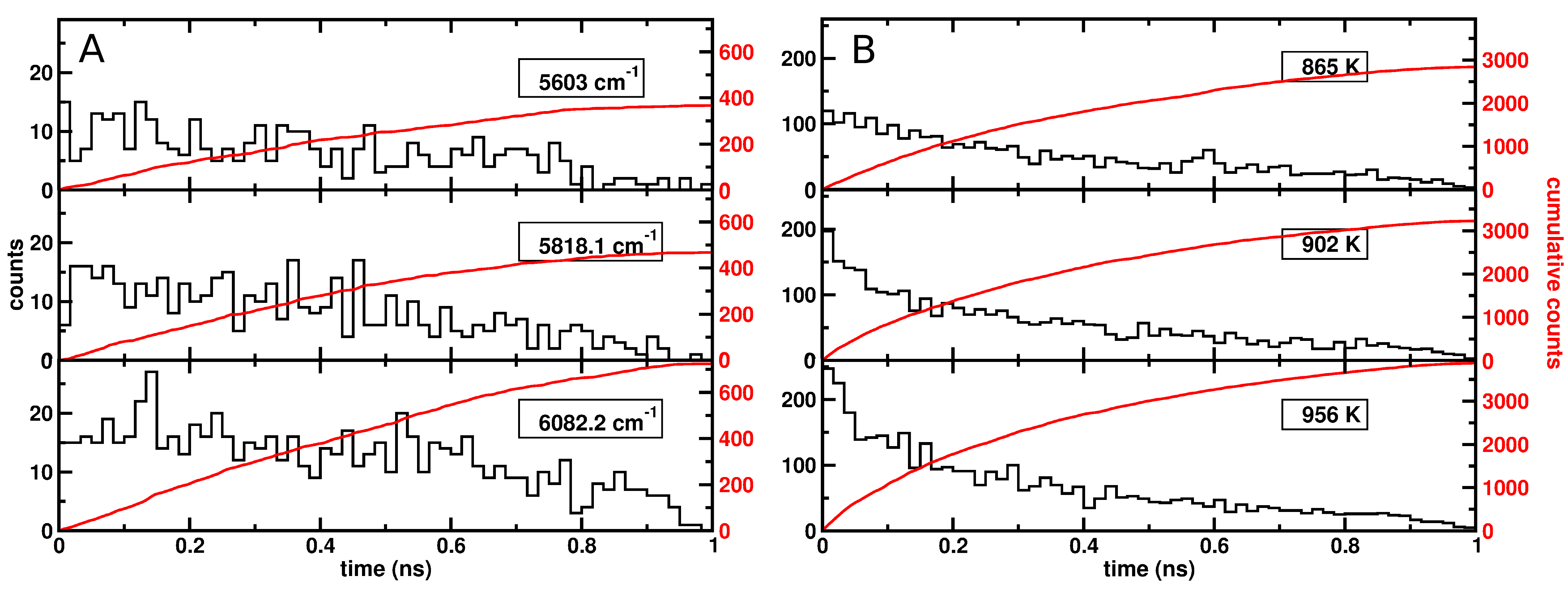}
    \caption{Distribution of vinyl hydroperoxide lifetimes before OH
      formation from vibEX (panel A) and thermal (panel B) using
      MS-ARMD and $D_e^{\rm OO} = 31.5$ kcal/mol. Histograms (black)
      and cumulative (red line) events are reported in the same panel
      with corresponding labels along the $y-$axes. Note the different
      scaling of the $y-$axes in panels A and B. The kinetic
      temperatures used to label the thermal simulations (panel B) are
      close to the excitation energies used in the vibEX simulations
      (panel A).}
    \label{sifig:vhp}
\end{figure}

\begin{figure}[h!]
  \centering \includegraphics[scale=0.48]{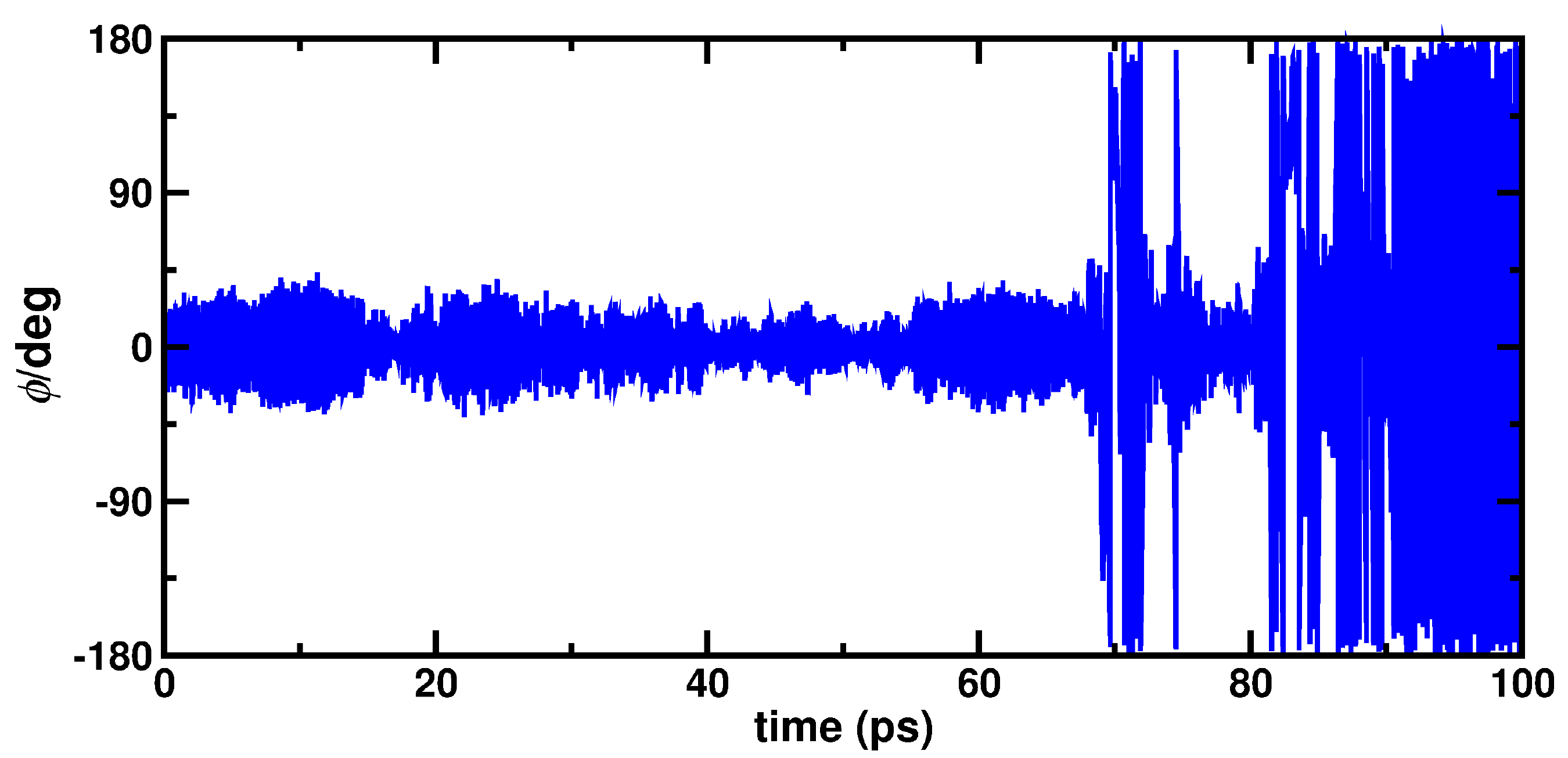}
 \caption{Time series $\phi(t)$ from a thermal trajectory at 902 K
   using ASE. Until $t = 67$ ps the system is in the
   \textit{syn-}CH$_3$CHOO state with $\phi \sim 0$. Between $t = 67$
   and 90 ps it is in VHP. Here, $\phi \in [-180.. +180]$, but most
   time is spent in the syn-form $\phi \in [-60.. +60]$. At $t= 90$ ps
   OH dissociation takes place and $\phi$ looses its meaning.}
\label{sifig:dihedral}
\end{figure}

\begin{figure}[h!]
   \centering \includegraphics[scale=0.30]{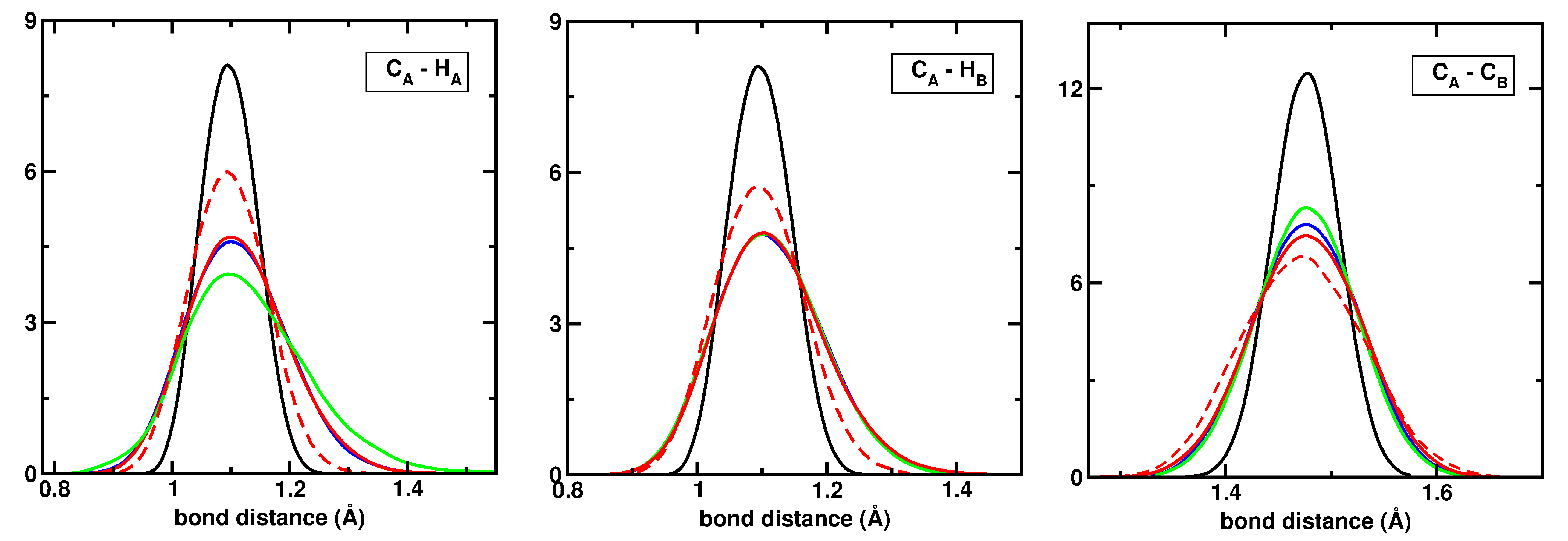}
  \caption{Bond distance distributions of C$_{\rm A}$H$_{\rm A}$,
    C$_{\rm A}$H$_{\rm B}$ and C$_{\rm A}$C$_{\rm B}$ for
    \textit{syn-}CH$_3$CHOO at 300 K (black) 250 ps, vibEX at 5983
    cm$^{-1}$ from 250 ps long simulation (blue) using the reactant
    FF, thermal simulations at 760 K from 250 ps long simulation (red)
    using the reactant FF and vibEX at 5983 cm$^{-1}$ from reactive
    simulation showing H-transfer within 10 ps (green) using
    MS-ARMD. Bond distance distribution from thermal simulations (red
    dashed) at 932 K using ASE.}
    \label{sifig:pofr1}
\end{figure}

\begin{figure}[h!]
   \centering
   \includegraphics[scale=0.32]{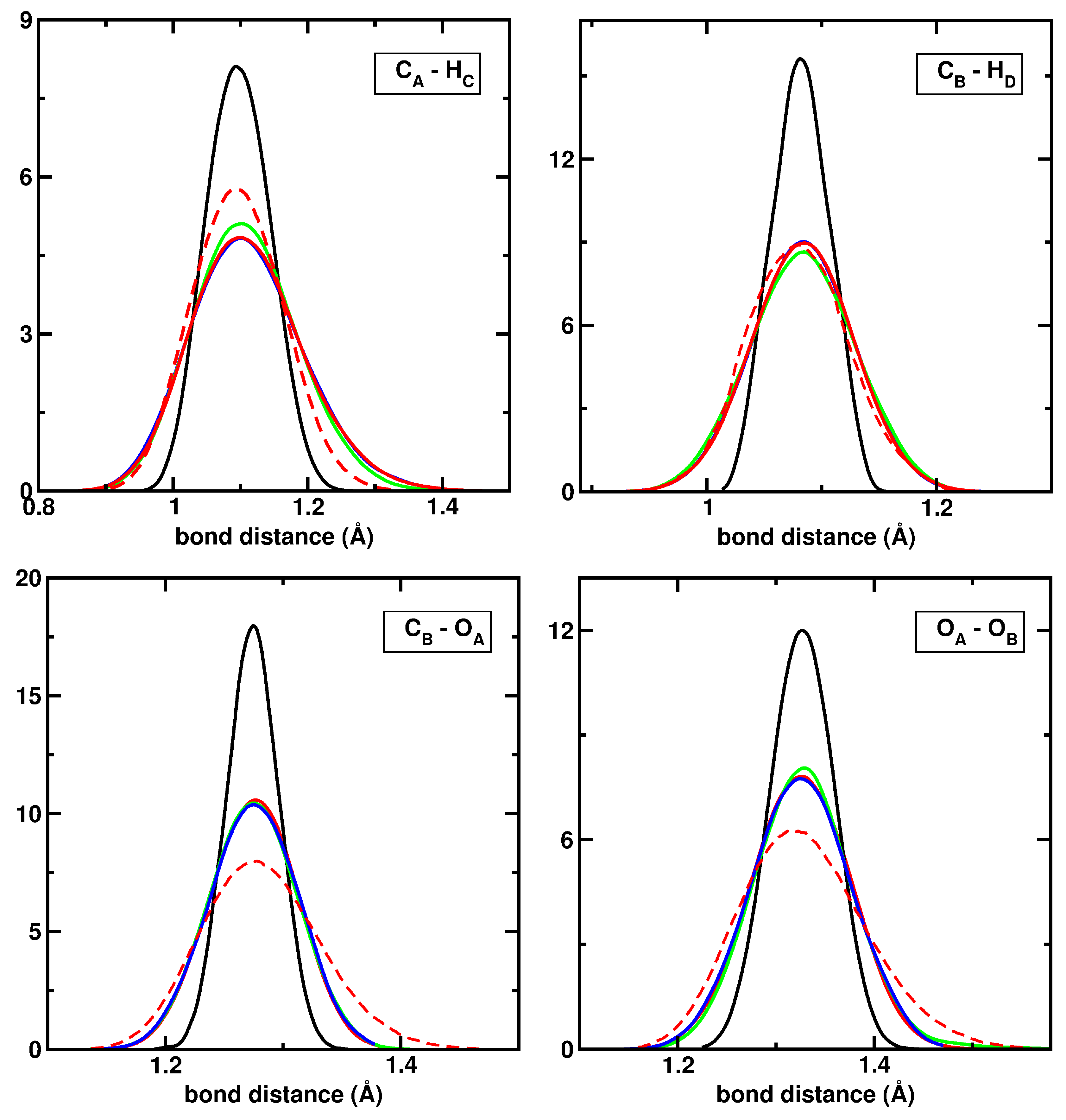}
  \caption{Bond distance distribution for {\it syn-}CH$_3$CHOO at 300
    K (black solid line), Vibrationally activated at 5983 cm$^{-1}$
    from 250 ps long simulation (blue solid line) using reactant FF,
    thermally activated at 760 K from 250 ps long simulation (red
    solid line) using reactant FF and Vibrationally activated at 5983
    cm$^{-1}$ from reactive simulation showing H-transfer in 10 ps
    (green solid line) using MS-ARMD. Bond distance distribution from
    thermal simulations (red dashed) at 932 K using ASE.}
    \label{sifig:pofr2}
\end{figure}

 \begin{table}[!ht]
     \centering
     \vspace{0.2cm}
\begin{tabular} { |P{2.88cm} |  P{0.85cm} | P{1.4cm} | P{1.38cm} ||   P{2.88cm} | P{0.85cm} | P{1.4cm} | P{1.38cm}  | }
   \hline IR excitation(cm$^{-1}$) & $t$ (ns) & H transfer & OH formed
   & IR excitation(cm$^{-1}$)& $t$ (ns) & H transfer & OH
   formed\\
   \hline
5603   & 0.1 & 1307 & 8    &  5950.9   &  0.1 &1860   & 12  \\
        & 0.2 & 2479 & 21   &           &  0.2 & 3286  & 42 \\
        & 0.3 & 3498 & 48   &           &  0.3 & 4484  &83 \\
        & 0.4 & 4378 & 85   &           &  0.4 & 5447  &147 \\
        & 0.5 & 5104 & 124  &           &  0.5 & 6272  &209 \\
        & 0.6 & 5679 & 158  &           &  0.6 & 6874  &267  \\
        & 0.7 & 6212 & 207  &           &  0.7 & 7344   &342 \\
        & 0.8 & 6700 & 254  &           &  0.8 & 7763   &422 \\
        & 0.9 & 7120 & 305  &           &  0.9 & 8089   &506 \\
        & 1.0 & 7452 & 367  &           &  1.0 & 8413   &605 \\

\hline
  5709 & 0.1 & 1462 & 8    &       5983.5 &  0.1 &1873  & 15  \\
      & 0.2 & 2641 & 24   &              & 0.2 & 3371  & 50 \\
      & 0.3 & 3709 & 49   &              & 0.3 & 4551  & 98 \\
      & 0.4 & 4568 & 81   &              & 0.4 & 5487  & 161 \\
      & 0.5 & 5325 & 134  &              & 0.5 & 6255  & 226 \\
      & 0.6 & 5938 & 188  &              & 0.6 & 6932  & 307  \\
      & 0.7 & 6474 & 253  &              & 0.7 & 7440  & 388 \\
      & 0.8 & 6953 & 316  &              & 0.8 & 7896  & 460 \\
      & 0.9 & 7346 & 376  &              & 0.9 & 8228  & 555 \\
      & 1.0 & 7712 & 431  &              & 1.0 & 8528  & 647 \\
    \hline
    5747.6  & 0.1 & 1544 & 10  & 6082.2 &  0.1 & 2128  & 15 \\
        & 0.2 & 2761 & 35  &        &  0.2 & 3664  & 50 \\
        & 0.3 & 3786 & 71  &        &  0.3 & 4825  & 101 \\
        & 0.4 & 4761 & 109 &        &  0.4 & 5853  & 160 \\
        & 0.5 & 5510 & 154 &        &  0.5 & 6661  & 232 \\
        & 0.6 & 6200 & 208 &        &  0.6 & 7295  & 311 \\
        & 0.7 & 6697 & 264 &        &  0.7 & 7788  & 406 \\
        & 0.8 & 7142 & 334 &        &  0.8 & 8225  & 509 \\
        & 0.9 & 7526 & 401 &        &  0.9 & 8547  & 612 \\
        & 1.0 & 7873 & 485 &        &  1.0 & 8774  & 724 \\   
    \hline   
5818.1   & 0.1 & 1695  & 9  &&& &\\
        & 0.2 & 2996 & 31 &&&&\\
         & 0.3 & 4119 & 73 &&&&\\
         & 0.4 & 4999 & 116 &&&&\\
         & 0.5 & 5782 & 169 &&&&\\
         & 0.6 & 6415 & 225 &&&& \\
         & 0.7 & 6943 & 276&&& &\\
         & 0.8 & 7381  & 340 &&&&\\
         & 0.9 & 7747  & 410&&& &\\
         & 1.0 & 8097 & 467&&& &\\
    \hline   
\end{tabular}
\vspace{0.2cm}
\caption{Total number of H- transfer and OH-elimination reactions from
  MS-ARMD vibEX simulations with $D_e^{\rm OO} = 31.5$ kcal/mol. The
  total simulation time is 1 ns for each trajectory and 10000
  independent simulations are carried out. The aggregated number of
  reactions up to time $t$ is reported.}
\label{sitab:armd31vib}
\end{table}

  \begin{table}[!ht]
     \centering
     \vspace{0.2cm}
   \begin{tabular} {| P{2.88cm} |  P{0.85cm} | P{1.4cm} | P{1.38cm} | | P{2.88cm} | P{0.85cm} | P{1.4cm} | P{1.38cm}  | } 

   \hline
   IR excitation(cm$^{-1}$) & $t$ (ns) & H transfer & OH formed & IR excitation(cm$^{-1}$)& $t$ (ns) & H transfer & OH formed\\
     
    \hline
    5603   & 10 & 242  &2  &5950.9& 10 & 267 & 3  \\      
        & 20 & 396  &7  &      & 20 & 460 & 17 \\
        & 30 & 534  &16 &      & 30 & 644 &27 \\
        & 40 & 674  &28 &      & 40 & 829 &52 \\
        & 50 & 795  &50 &      & 50 & 1043& 74 \\ 
        & 60 & 935  &68 &      & 60 & 1242&105  \\ 
        & 70 & 1084 &101&      & 70 & 1409&133 \\ 
        & 80 & 1215 &128&      & 80 & 1579&174 \\ 
        & 90 & 1327 &153&      & 90 & 1749&215 \\ 
        & 100 &1440 &178&      & 100 & 1927&257 \\ 
    \hline
   5709 &  10 & 215   &3   &5983.5&  10 &  311  &5  \\
     &  20 & 361   &12  &      &  20 &  513  &23 \\
     &  30 & 512   &19  &      &  30 &  706  &37 \\
     &  40 & 655   &27  &      &  40 &  881  &58 \\
     &  50 & 814   &54  &      &  50 &  1053 &84 \\
     &  60 & 942   &74  &      &  60 &  1264 &118  \\
     &  70 & 1090  &90  &      &  70 &  1440 &154 \\
     &  80 & 1232  &116 &      &  80 &  1599 &181 \\
     &  90 & 1351  &144 &      &  90 &  1803 &227 \\
     &  100& 1467  &178 &      &  100 & 1956 &275 \\

    \hline
5747.6   & 10 &  238 &3    &6082.2   & 10 & 337 & 4  \\
         & 20 &  409 &11   &         & 20 & 556 & 18 \\
         & 30 &  585 &21   &         & 30 & 766 & 38 \\
         & 40 &  753 &41   &         & 40 & 978 & 67 \\
         & 50 &  912 &62   &         & 50 & 1195 & 108 \\
         & 60 &  1060 &90  &         & 60 & 1414 & 140\\
         & 70 &  1187 &110 &         & 70 & 1628 & 182 \\
         & 80 &  1328 &145 &         & 80 & 1816 & 226 \\
         & 90 &  1458 &178 &         & 90 & 1990 & 288 \\
         & 100 & 1578 &212 &         & 100 & 2176 & 344 \\

    \hline   
5818.1   & 10 & 262& 3 &&&& \\
         & 20 & 450&12 &&&&\\
         & 30 &607 &28 &&&&\\
         & 40 & 791&48 &&&&\\
         & 50 &946 &68 &&&&\\
         & 60 &1104 &95 &&&& \\
         & 70 &1244 &132 &&&&\\
         & 80 &1390 &172 &&&&\\
         & 90 &1562 &218&&&& \\
         & 100 &1711 &253&&&&\\  
    \hline

\end{tabular}
\vspace{0.2cm}
\caption{Total number of H- transfer and OH-elimination reactions from
  MS-ARMD vibEX simulations with $D_e^{\rm OO} = 23.5$ kcal/mol. The
  total simulation time is 100 ps for each trajectory and 10000
  independent simulations are carried out. The aggregated number of
  reactions up to time $t$ is reported.}
\label{sitab:armd23vib}
\end{table}   
  
   \begin{table}[!ht]
     \centering
     \vspace{0.2cm}
    
\begin{tabular} { |P{1.5cm} |  P{0.85cm} | P{0.85cm} | P{1.4cm} |   P{1.4cm} || P{1.5cm} | P{0.85cm} | P{0.85cm}| P{1.4cm} |P{1.4cm} | }
   \hline
     $T$ (K) & Total & $t$ (ns) & H transfer & OH formed & $T$ (K) & Total & $t$ (ns) & H transfer & OH formed \\
    \hline
  865 & 9292 & 0.1 & 4034 & 416  &   925 & 8894 & 0.1 & 4648 & 657 \\
        & & 0.2 & 5272 & 861     &       & & 0.2 &  5816 & 1230 \\
        & & 0.3 & 5905 & 1236    &       & & 0.3 &  6395 & 1668 \\
        & & 0.4 & 6328 & 1561    &       & & 0.4 &  6764 &2094 \\
        & & 0.5 & 6642 & 1831    &       & & 0.5 &  7006 &2414 \\
        & & 0.6 & 6858 & 2076    &       & & 0.6 &  7193 &2685  \\
        & & 0.7 & 7039 & 2295    &       & & 0.7 & 7337  &2897 \\
        & & 0.8 & 7172 & 2501    &       & & 0.8 & 7466  &3117 \\
        & & 0.9 & 7268 & 2680    &       & & 0.9 & 7551  &3294 \\
        & & 1.0 & 7377 & 2839    &       & & 1.0 & 7621  &3449 \\

    \hline
   877 & 9227 & 0.1 & 4228 & 510   &   932 & 8848 & 0.1 & 4663 & 690  \\
        & & 0.2 & 5351 & 971     &         & & 0.2 &  5777 &1311 \\
        & & 0.3 & 5959 & 1336    &         & & 0.3 &  6358 &1798\\
        & & 0.4 & 6356 & 1684    &         & & 0.4 &  6687 & 2178 \\
        & & 0.5 & 6640 & 1953    &         & & 0.5 &  6960 &2472 \\
        & & 0.6 & 6875 & 2181    &         & & 0.6 &  7138 &2746  \\
        & & 0.7 & 7048 & 2442    &         & & 0.7 & 7305  &2983 \\
        & & 0.8 & 7184 & 2610    &         & & 0.8 & 7406  &3165 \\
        & & 0.9 & 7308 & 2771    &         & & 0.9 & 7520  & 3364 \\
        & & 1.0 & 7419 & 2924    &         & & 1.0 & 7590  & 3525 \\

    \hline

886  & 9165 & 0.1 & 4269 &490 &956 & 8789 & 0.1 & 4982 &827 \\
        & & 0.2 & 5415 &974   &        & & 0.2 & 6076  &1532 \\
        & & 0.3 & 6054 & 1374 &        & & 0.3 & 6620  &2026 \\
        & & 0.4 & 6428 & 1717 &        & & 0.4 & 6948 &2455 \\
        & & 0.5 & 6708 & 1981 &        & & 0.5 & 7152  &2774 \\
        & & 0.6 & 6922  & 2273&        & & 0.6 & 7319 &3063 \\
        & & 0.7 & 7069  & 2477&        & & 0.7 & 7435 &3308 \\
        & & 0.8 & 7207  & 2701&        & & 0.8 & 7535  &3528 \\
        & & 0.9 & 7339 & 2871 &        & & 0.9 & 7621  &3717 \\
        & & 1.0 & 7449 & 3039 &        & & 1.0 & 7683 &3880 \\

    \hline   
902  & 9032 & 0.1 & 4347  & 601 &&&&& \\
        & & 0.2 & 5514 & 1115 &&&&&\\
        & & 0.3 & 6123 & 1539 &&&&&\\
        & & 0.4 & 6503 & 1895 &&&&&\\
        & & 0.5 & 6748 & 2191 &&&&&\\
        & & 0.6 & 6993 & 2454 &&&&& \\
        & & 0.7 & 7179 & 2684 &&&&&\\
        & & 0.8 & 7307  & 2876 &&&&&\\
        & & 0.9 & 7392  & 3068 &&&&&\\
        & & 1.0 & 7493 & 3228 &&&&&\\  
    \hline

\end{tabular}
\vspace{0.2cm}
\caption{Total number of H- transfer and OH-elimination reactions from
  MS-ARMD thermal simulations with $D_e^{\rm OO} = 31.5$ kcal/mol. The
  total simulation time is 1 ns for each trajectory and 10000
  independent simulations are carried out. The aggregated number of
  reactions up to time $t$ is reported. Some simulations shown
  H-transfer in heating step which are excluded from the total.}
\label{sitab:armd31th}
\end{table}

   \begin{table}[!ht]
     \centering
     \vspace{0.2cm}
    \begin{tabular} { |P{1.5cm} |  P{0.85cm} | P{0.85cm} | P{1.4cm} |   P{1.4cm} || P{1.5cm} | P{0.85cm} | P{0.85cm}| P{1.4cm} |P{1.4cm} | }
   \hline
     $T$ (K) & Total & $t$ (ns) & H transfer & OH formed & $T$ (K) & Total & $t$ (ns) & H transfer & OH formed \\

    \hline
 865  &9338 & 10 & 827 &93  &           925 &9000 & 10 & 1239 &204   \\
        & & 20 & 1496 & 286    &               & & 20 &  2103 & 542\\
        & & 30 & 2022 & 524    &               & & 30 & 2682 & 905\\
        & & 40 & 2426 & 768    &               & & 40 & 3143 & 1249\\
        & & 50 & 2765 & 998    &               & & 50 &  3538 & 1533\\
        & & 60 & 3081 & 1224   &               & & 60 &  3846 & 1818 \\
        & & 70 & 3332 & 1440   &               & & 70 &  4139 & 2095\\
        & & 80 & 3554 & 1631   &               & & 80 & 4370  & 2342\\
        & & 90 & 3767 & 1827   &               & & 90 & 4574  & 2555\\
        & & 100 & 3970 & 2023   &              & & 100 &  4752 & 2768 \\

    \hline
  877 &9324 & 10 & 954 & 146    &   932   &9000 & 10 &  1231 & 170 \\
        & & 20 & 1651 & 385    &           & & 20 & 2112  & 472\\
        & & 30 & 2196 & 615    &           & & 30 & 2721  & 835\\
        & & 40 & 2604 & 860    &           & & 40 &  3203 & 1196 \\
        & & 50 & 2989 & 1107   &           & & 50 &  3599 & 1525\\
        & & 60 & 3279 & 1352   &           & & 60 &  3911  & 1823  \\
        & & 70 & 3560 & 1575   &           & & 70 &  4167 & 2100\\
        & & 80 & 3787 & 1788   &           & & 80 &  4396 & 2349\\
        & & 90 & 4016 & 1973   &           & & 90 &  4563 & 2550\\
        & & 100 & 4188 & 2173  &           & & 100 & 4750  & 2770\\

    \hline
886  & 9257 & 1037 & 34 & 133      &   956  &8917 & 10 &  1356 & 216 \\
        & & 20 & 1768 & 401        &           & & 20 & 2268  & 616\\
        & & 30 & 2283 & 650        &           & & 30 & 2899  & 997\\
        & & 40 & 2698 & 922        &           & & 40 &  3382 & 1357\\
        & & 50 & 3103 & 1174       &           & & 50 &  3753 & 1695\\
        & & 60 & 3411  & 1406      &           & & 60 &  4085  & 2011  \\
        & & 70 & 3687  & 1642      &           & & 70 &  4373 & 2277\\
        & & 80 & 3900 & 1850       &           & & 80 &  4623 & 2560\\
        & & 90 &  4130 & 2066      &           & & 90 &  4794 & 2795\\
        & & 100 & 4332 & 2258      &           & & 100 & 4989  & 3014\\
    \hline   
902 & 9167 & 10 & 1065  &  148&&&&&\\
       &  & 20 & 1822 &  410&&&&&\\
        & & 30 & 2427  & 681 &&&&&\\
        & & 40 & 2876  & 986 &&&&&\\
        & & 50 & 3277  & 1232 &&&&&\\
        & & 60 &  3601 &  1520 &&&&&\\
        & & 70 &  3881 &  1766&&&&&\\
        & & 80 &  4113 & 2001 &&&&&\\
        & & 90 & 4334  & 2232 &&&&&\\
        & & 100 &4538 & 2407 &&&&&\\  
    \hline

\end{tabular}
\vspace{0.2cm}
\caption{Total number of H- transfer and OH-elimination reactions from
  MS-ARMD thermal simulations with $D_e^{\rm OO} = 23.5$ kcal/mol. The
  total simulation time is 100 ps for each trajectory and 10000
  independent simulations are carried out. The aggregated number of
  reactions up to time $t$ is reported. Some simulations shown
  H-transfer in heating step which are excluded from the total.}
\label{sitab:armd23th}
\end{table}

     \begin{table}[!ht]
     \centering
     \vspace{0.2cm}
     \begin{tabular} { |P{1.5cm} |  P{0.85cm} | P{0.85cm} | P{1.4cm} |   P{1.4cm} | | P{1.5cm} | P{0.85cm} | P{0.85cm}| P{1.4cm} |P{1.4cm} | }
   \hline
     $T$ (K) & Total & $t$ (ns) & H transfer & OH formed & $T$ (K) & Total & $t$ (ns) & H transfer & OH formed \\

    \hline
  865 &9980 & 10 &15 &  12 & 902 & 9891 & 10 & 30 & 24  \\
          & & 20 &35 &  29 & & &          20 &48 & 43\\
          & & 30 &44 &  43 & & &          30 &68 & 62\\
          & & 40 & 56&  54 & & &          40&90 & 84\\
          & & 50 & 69&  66 & & &          50&108 & 103\\
          & & 60 &78 &  73 & & &          60&121 & 116\\
          & & 70 &90 &  88 & & &          70& 134& 131\\
          & & 80 &99 &  98 & & &          80&144 & 143\\
          & & 90 &107 &  106& & &          90& 161& 153\\
          & & 100 &112 & 110& & &         100&171 & 169\\
    \hline
   877 &9978 & 10 & 24 & 19 & 925 & 9970 & 10 &40 & 30 \\
        & & 20 &38 & 34   & & &          20 & 58& 53\\
        & & 30 &51 & 48   & & &          30 & 75& 70\\
        & & 40 &62 & 62   & & &          40 & 100& 92 \\
        & & 50 & 74& 74   & & &          50 & 117& 113 \\
        & & 60 & 90& 83   & & &          60 & 141& 136\\
        & & 70 & 102& 97   & & &          70 &165 & 159\\
        & & 80 & 120& 114  & & &          80 &175 & 172\\
        & & 90 & 131& 130  & & &          90 &195 & 188\\
        & & 100 & 141& 139 & & &          100 &211 & 208\\  
    \hline
886  & 9956& 10 &26 &21 & 956 & 9961 & 10 &41 & 34  \\
      &   & 20 &48& 42 & & &          20 &77 & 69\\
       &  & 30 &62& 55 & & &          30 &108 & 102\\
    &     & 40 &85& 75 & & &          40 & 144& 133\\
     &    & 50 &99& 98 & & &          50 &164 & 162\\
    &     & 60 &112&  110 & & &        60 &192 & 180\\
    &     & 70 &126& 125 & & &          70 &208 & 207\\
    &     & 80 &143& 139 & & &          80 & 227& 225\\
    &     & 90 &153& 149 & & &          90 & 242& 239\\
    &     & 100 &164& 162 & & &         100 &260 & 260\\  
    \hline   
\end{tabular}
\vspace{0.2cm}
\caption{Total number of H- transfer and
  OH-elimination reactions from PhysNet thermal simulations. The total simulation time is 100 ps
  for each trajectory and 10000 independent simulations are carried
  out. The aggregated number of reactions up to time $t$ is reported. Some simulations shown H-transfer in heating step which are excluded from the total.}
\label{sitab:aseth}
\end{table} 
  
 \clearpage
\bibliography{refs}